\documentclass[twocolumn,showpacs,amsmath,amssymb,aps,prb]{revtex4}

\usepackage{graphicx}
\usepackage{multirow}
\usepackage{color}
\usepackage{dcolumn}
\usepackage{bm}
\usepackage{subfigure}
\newcommand{\bra}[1] {\langle #1 |}
\newcommand{\ket}[1] {| #1 \rangle}

\newcommand{\ketbra}[2] {| #1 \rangle\langle #2 |}

\newcommand{\rdm}[3]{\bra{#1}| #2 | \ket{#3}}

\begin{document}

\title{Theory of the ground state spin of the NV$^-$ center in diamond: II. Spin solutions, time-evolution, relaxation and inhomogeneous dephasing}

\author{M.W. Doherty$^1$, F. Dolde$^2$, H. Fedder$^2$, F. Jelezko$^{2,3}$, J. Wrachtrup$^2$, N.B. Manson$^4$, and L.C.L. Hollenberg$^1$}

\affiliation{$^1$ School of Physics, University of Melbourne, Victoria 3010, Australia \\
$^2$ 3$^{\mathrm{rd}}$ Institute of Physics and Research Center SCOPE, University Stuttgart, Pfaffenwaldring 57, D-70550 Stuttgart, Germany \\
$^3$ Institut f$\mathrm{\ddot{u}}$r Quantenoptik, Universit$\mathrm{\ddot{a}}$t Ulm, Ulm D-89073, Germany \\
$^4$ Laser Physics Centre, Research School of Physics and Engineering, Australian National University, Australian Capital Territory 0200, Australia}

\date{11 November 2011}

\begin{abstract}
The ground state spin of the negatively charged nitrogen-vacancy center in diamond has many exciting applications in quantum metrology and solid state quantum information processing, including magnetometry, electrometry, quantum memory and quantum optical networks. Each of these applications involve the interaction of the spin with some configuration of electric, magnetic and strain fields, however, to date there does not exist a detailed model of the spin's interactions with such fields, nor an understanding of how the fields influence the time-evolution of the spin and its relaxation and inhomogeneous dephasing. In this work, a general solution is obtained for the spin in any given electric-magnetic-strain field configuration for the first time, and the influence of the fields on the evolution of the spin is examined. Thus, this work provides the essential theoretical tools for the precise control and modeling of this remarkable spin in its current and future applications.
\end{abstract}


\maketitle

\section{Introduction}

The negatively charged nitrogen-vacancy (NV$^-$) center in diamond has many exceptional properties that are highly suited to applications in quantum metrology and quantum information processing (QIP). The exciting recent demonstrations of high precision magnetometry,\cite{mag1,mag2,mag3,mag4,mag5,mag6,coherence} electrometry,\cite{efield} decoherence based imaging,\cite{decoherence1,decoherence2,decoherence3,decoherence4,decoherence5} and spin-photon \cite{spinphoton} and spin-spin \cite{spinspin1,spinspin2,spinspin3,spinspin4,spinspin5} entanglement have each utilized the ground state spin as a solid state spin qubit. Indeed, these demonstrations exploit the interaction of the spin with some configuration of electric, magnetic and strain fields and the center's remarkable capability of optical spin-polarization and readout. \cite{readout1,readout2} The theory of the ground state spin was presented in part I of this paper series, \cite{partI} including the spin's fine and hyperfine structures, and its interactions with the different fields. The theory will now be applied to derive the solutions and dynamics of the spin for a general field configuration, including the effects of field inhomogeneities and the influence of fields on the spin's relaxation.

\begin{figure}[hbtp]
\begin{center}
\includegraphics[width=0.8\columnwidth] {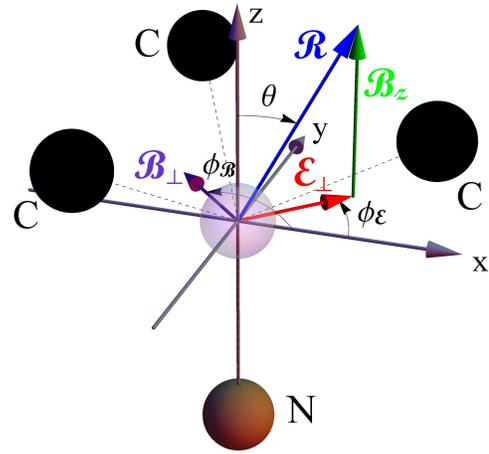}
\caption{(color online) Schematic of the nitrogen-vacancy center and the adopted coordinate system, depicting: the vacancy (transparent), the nearest-neighbor carbon atoms to the vacancy (black), the substitutional nitrogen atom (brown), the effective magnetic and electric-strain field components (colored arrows), and their corresponding field angles.}
\label{fig:centre}
\end{center}
\end{figure}

\begin{figure}[hbtp]
\begin{center}
\includegraphics[width=1.0\columnwidth] {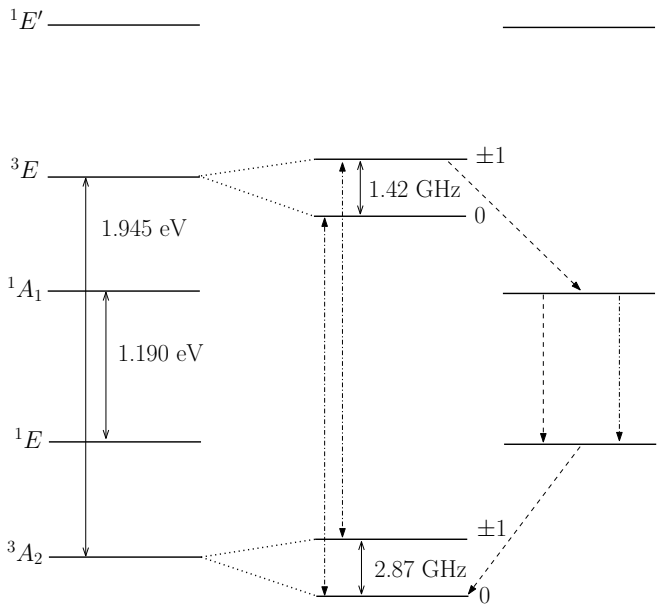}
\caption{The electronic orbital structure (left) and fine structure (right) of the NV$^-$ center at ambient temperatures. The observed optical zero phonon line (1.945 eV) \cite{dupreez} and infrared zero phonon line (1.190 eV) \cite{infrared} transitions are depicted as solid arrows in the orbital structure. The radiative (chain arrows) and non-radiative (dashed arrows) pathways that result in the optical spin-polarization and readout of the center are depicted in the fine structure. Note that the much weaker radiative and non-radiative transitions that act to reduce spin-polarization have not been depicted.}
\label{fig:electronicstructure}\label{fig:spinpol}
\end{center}
\end{figure}

The NV$^-$ center is a point defect of $C_{3v}$ symmetry in diamond consisting of a substitutional nitrogen atom adjacent to a carbon vacancy (refer to Fig. \ref{fig:centre}). The center's electronic structure is summarized in Fig. \ref{fig:electronicstructure}. It consists of a $^3A_2$ ground triplet state, an optical $^3E$ excited triplet and several dark singlet states. \cite{njp} At ambient temperatures, the fine structures of the ground and excited triplet states have single zero-field splittings between the $m_s = 0$ and $m_s = \pm1$ spin sub-levels of $D_{gs}\sim 2.87$ GHz  and $D_{es} \sim 1.42$ GHz respectively. \cite{njp} Zeeman, Stark and strain splittings have been observed in the fine structures of both triplet states, \cite{excitedstatestrain,excitedstatezeeman,ground,tamarat,vanoort} although the Stark and strain effects in the ground triplet state are several orders of magnitude smaller than in the excited triplet state. \cite{vanoort}

One of the most intriguing properties of the NV$^-$ center is the ability to optically polarize and readout the ground state spin, \cite{readout1,readout2} distinguishing the center from other paramagnetic defects in diamond, and forming the basis for its QIP and quantum metrology applications. As depicted in Fig. \ref{fig:spinpol}, the process of optical spin-polarization occurs due to the presence of a non-radiative decay pathway from the excited triplet state to the ground triplet state that competes with the optical decay pathway. The details of the photokinetics of the non-radiative pathway are yet to be fully explained,  but it is believed that the $m_s = \pm1$ sub-levels of the excited triplet state are preferentially depopulated and the $m_s = 0$ sub-level of the ground triplet state is preferentially populated, thereby polarizing the population into the $m_s =0$ spin state after optical cycling. \cite{readout1,readout2} The preferential non-radiative depopulation of the $m_s = \pm1$ sub-levels of the excited triplet state also introduces a difference in the optical emission intensity between the spin sub-levels. This difference in emission intensity can be utilized to readout the relative populations of the $m_s = 0$ and $m_s = \pm1$ sub-levels of the ground triplet state through the measurement of the integrated emission intensity upon optical excitation. \cite{readout1,readout2}

The center's capability of optical spin-polarization and readout enables the implementation of continuous wave and pulsed optically detected magnetic resonance (ODMR) techniques. \cite{readout1,readout2} Simple pulsed ODMR techniques such as free induction decay (FID) and spin echo\cite{spinecho1,spinecho2} as well as more complicated multi-pulse ODMR techniques\cite{mag6,multipulse1,multipulse2} have been implemented in the center's quantum metrology and QIP applications and involve optical polarization and readout pulses encompassing a sequence of microwave pulses tuned to the fine structure splittings of the ground triplet state. The microwave pulses coherently manipulate the ground state spin and result in an optically detectable oscillation in the relative population of the $m_s = 0$ and $m_s = \pm1$ sub-levels. In order to optimally control the spin and illicit the maximum amount of information and sensitivity from its implementation in ODMR experiments, a detailed model of the time-evolution, relaxation and dephasing of the spin is required.

In this article, the detailed theoretical construction of the ground state spin-Hamiltonian that was conducted in Part I\cite{partI} of this paper series will be applied in order to produce the solution of the ground state spin and its time-evolution in the presence of a general electric-magnetic-strain field configuration. The solution will be demonstrated by modeling a simple Free Induction Decay (FID) experiment and examining the dependence of the FID signal, inhomogeneous dephasing and spin relaxation on the applied field configuration. This simple demonstration will provide insight into the observed strong dependence of the spin's inhomogeneous dephasing time on the applied fields. \cite{efield} Furthermore, the spin solution will be a useful tool in future applications of the spin in quantum metrology and QIP as it clearly describes how electric, magnetic and strain fields can be used to precisely control this important spin in diamond. For example, multimodal decoherence microscopy that maps both magnetic and electric noise using the same probe.\cite{decoherence1,decoherence2,decoherence3,decoherence4,decoherence5}

\section{Solutions of the ground state spin}

The ground state spin-Hamiltonian as derived in Part I is
\begin{eqnarray}
\hat{H}_{gs} &=& \frac{1}{\hbar^2}{\cal D}S_z^2+\frac{1}{\hbar}\vec{S}\cdot\vec{{\cal B}}-\frac{1}{\hbar^2}{\cal E}_x(S_x^2-S_y^2) \nonumber \\
&&+\frac{1}{\hbar^2}{\cal E}_y(S_xS_y+S_yS_x)\label{eq:spinhamiltonian}
\end{eqnarray}
where $\vec{S} = S_x\hat{\vec{x}}+ S_y\hat{\vec{y}}+ S_z\hat{\vec{z}}$ is the total electronic spin operator for $S=1$, ${\cal D}$ is the effective spin-spin and axial electric-strain field, $\vec{{\cal B}}$ is the effective magnetic field, and ${\cal E}_x $ and ${\cal E}_y$ are the effective non-axial electric-strain field components. Refer to Part I  for the explicit definitions of the effective fields. As discussed in Part I, the solutions of $\hat{H}_{gs}$ describe both the interactions of the electronic spin-orbit states in the high field limit, where the field induced shifts are much larger than the spin's hyperfine structure, and also the interactions of the $m_I = 0$ subset of hyperfine states in the weak field limit, where the field induced shifts are comparable to the spin's hyperfine structure.

It is convenient to define the field spin states $\{\ket{0},\ket{-},\ket{+}\}$ in terms of the $S_z$ eigenstates $\{\ket{S,m_s}\}$ as
\begin{eqnarray}
&&\ket{0} = \ket{1,0} \nonumber \\
&&\ket{-} = e^{i\frac{\phi_{\cal E}}{2}}\sin\frac{\theta}{2}\ket{1,1}+e^{-i\frac{\phi_{\cal E}}{2}}\cos\frac{\theta}{2}\ket{1,-1}, \nonumber \\
&&\ket{+} = e^{i\frac{\phi_{\cal E}}{2}}\cos\frac{\theta}{2}\ket{1,1}-e^{-i\frac{\phi_{\cal E}}{2}}\sin\frac{\theta}{2}\ket{1,-1}
\end{eqnarray}
where $\tan\phi_{{\cal E}}={\cal E}_y/{\cal E}_x$, $\tan\theta = {\cal E}_\perp/{\cal B}_z$ and ${\cal E}_\perp = \sqrt{{\cal E}_x^2+{\cal E}_y^2}$. The matrix representation of $\hat{H}_{gs}$ in the basis $\{\ket{0},\ket{-},\ket{+}\}$ is
\begin{widetext}
\begin{eqnarray}
H_{gs} =
\left(\begin{array}{ccc}
0 & \frac{{\cal B}_\perp}{\sqrt{2}}\left(e^{i\frac{\phi}{2}}s_{\frac{\theta}{2}}+e^{-i\frac{\phi}{2}}c_{\frac{\theta}{2}}\right) & \frac{{\cal B}_\perp}{\sqrt{2}}\left(e^{i\frac{\phi}{2}}c_{\frac{\theta}{2}}-e^{-i\frac{\phi}{2}}s_{\frac{\theta}{2}}\right) \\
\frac{{\cal B}_\perp}{\sqrt{2}}\left(e^{-i\frac{\phi}{2}}s_{\frac{\theta}{2}}+e^{i\frac{\phi}{2}}c_{\frac{\theta}{2}}\right) & {\cal D}-{\cal R} & 0 \\
\frac{{\cal B}_\perp}{\sqrt{2}}\left(e^{-i\frac{\phi}{2}}c_{\frac{\theta}{2}}-e^{i\frac{\phi}{2}}s_{\frac{\theta}{2}}\right) & 0 & {\cal D}+{\cal R} \\
\end{array}\right) \nonumber \\ \label{eq:solspinhamiltonian}
\end{eqnarray}
\end{widetext}
where $\phi = 2\phi_{{\cal B}}+\phi_{{\cal E}}$, $\tan\phi_{{\cal B}} = {\cal B}_y/{\cal B}_x$, ${\cal R} = \sqrt{{\cal B}_z^2+{\cal E}_\perp^2}$, $c_{\frac{\theta}{2}} = \cos\frac{\theta}{2}$ and $s_{\frac{\theta}{2}} = \sin\frac{\theta}{2}$. Therefore, if ${\cal B}_\perp = \sqrt{{\cal B}_x^2+{\cal B}_y^2} \ll {\cal D}$ and ${\cal B}_\perp^2/ {\cal D} \ll {\cal R}$, the field spin states are approximate eigenstates of $\hat{H}_{gs}$ with energies $E_0 = 0$ and $E_\pm = {\cal D}\pm{\cal R}$. In this weak non-axial magnetic field limit, the spin eigenstates are completely characterized by the field angles $\phi_{\cal E}$ and $\theta$ derived from the axial magnetic and non-axial electric-strain field components and the energy splitting of the $\ket{\pm}$ states is governed by the magnitudes of the same field components.

When ${\cal B}_\perp^2/ {\cal D} \sim {\cal R}$, first-order corrections to the field spin states become important, such that
\begin{eqnarray}
\ket{0}^{(1)} & = &  \ket{0}-\sqrt{\frac{\Delta{\cal R}}{{\cal D}}}\left(e^{-i\frac{\phi}{2}}s_{\frac{\theta}{2}}+e^{i\frac{\phi}{2}}c_{\frac{\theta}{2}}\right)\ket{-} \nonumber \\
&& -\sqrt{\frac{\Delta{\cal R}}{{\cal D}}}\left(e^{-i\frac{\phi}{2}}c_{\frac{\theta}{2}}-e^{i\frac{\phi}{2}}s_{\frac{\theta}{2}}\right)\ket{+} \nonumber \\
\ket{-}^{(1)} & = & \ket{-}+\sqrt{\frac{\Delta{\cal R}}{{\cal D}}}\left(e^{i\frac{\phi}{2}}s_{\frac{\theta}{2}}+e^{-i\frac{\phi}{2}}c_{\frac{\theta}{2}}\right)\ket{0} \nonumber \\
\ket{+}^{(1)} & = & \ket{+}+\sqrt{\frac{\Delta{\cal R}}{{\cal D}}}\left(e^{i\frac{\phi}{2}}c_{\frac{\theta}{2}}-e^{-i\frac{\phi}{2}}s_{\frac{\theta}{2}}\right)\ket{0}
\end{eqnarray}
where $\Delta = {\cal B}_\perp^2/ 2{\cal R} {\cal D}$, and second-order perturbation corrections to the energies also become important, such that $E_0^{(2)} = -\Delta{\cal R}$ and
\begin{eqnarray}
E_\pm^{(2)} = {\cal D}+\Delta{\cal R}\pm{\cal R}\left(1-2\Delta\sin\theta\cos\phi+\Delta^2\right)^{\frac{1}{2}}
\end{eqnarray}
Consequently, in the strong non-axial magnetic field limit, the spin eigenstates and energies become dependent on the non-axial magnetic field direction $\phi_{\cal B}$ and the dimensionless ratio of the field magnitudes $\Delta$.

\begin{figure*}[hbtp]
\begin{center}
\mbox{
\subfigure[]{\includegraphics[width=0.8\columnwidth] {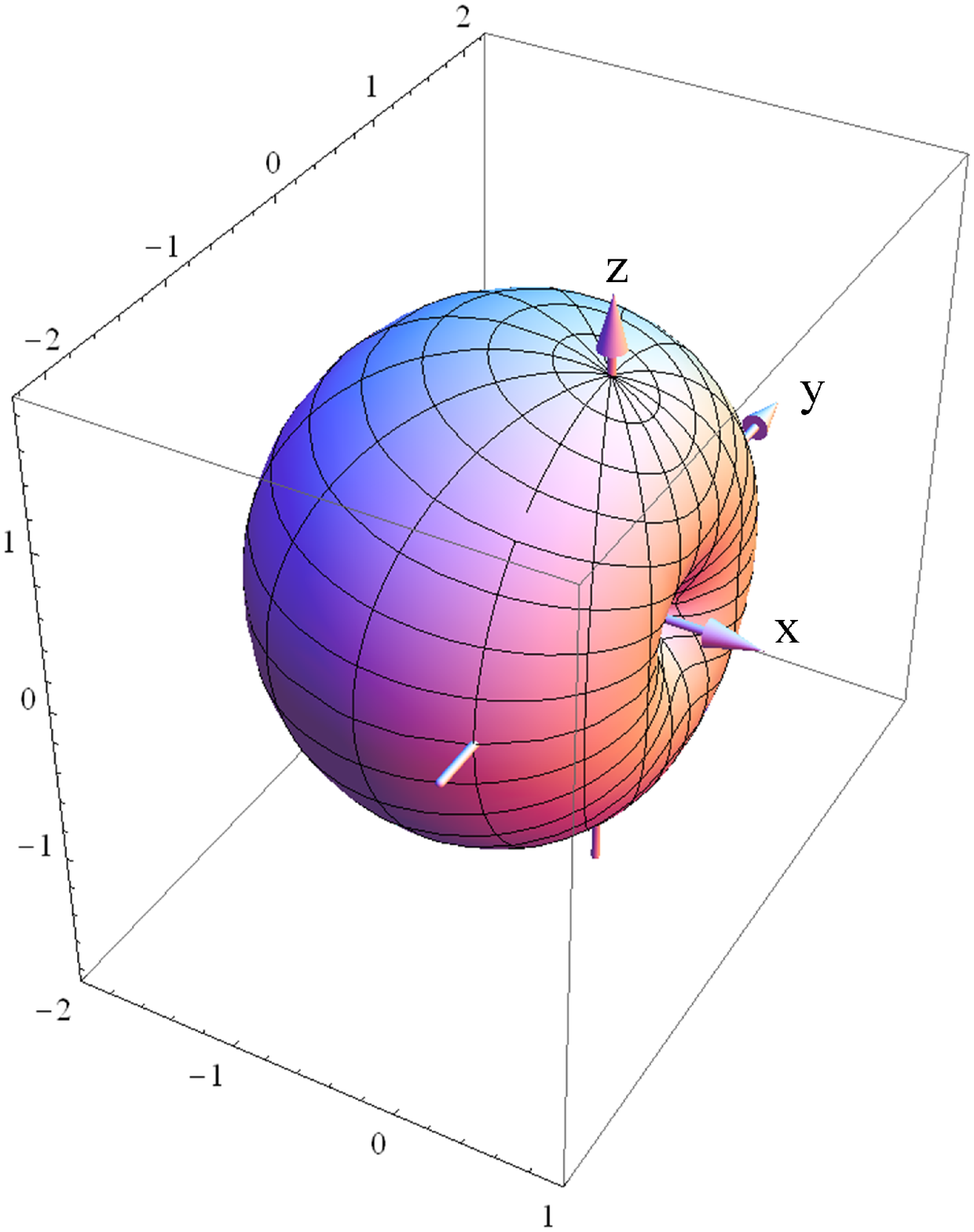}}
\subfigure[]{\includegraphics[width=0.65\columnwidth] {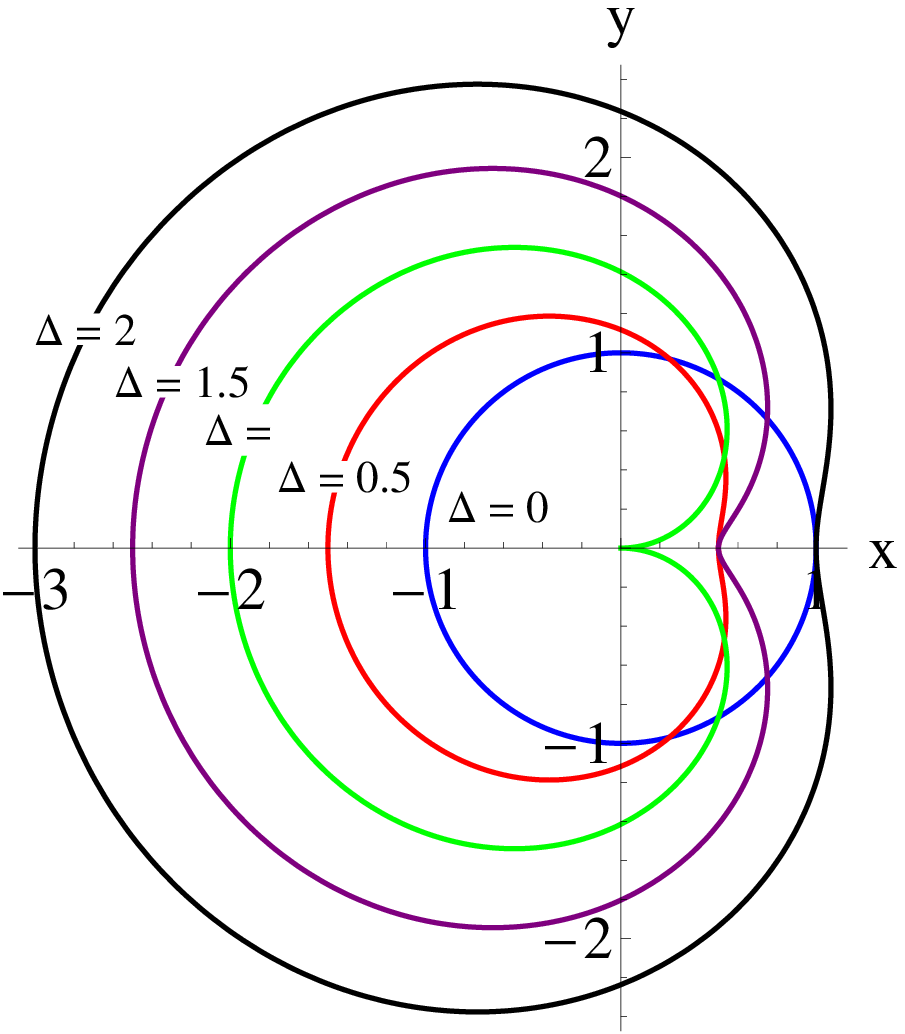}}
}
\caption{(color online) Plots of the dimensionless energy splitting parameter $\left(1-2\Delta\sin\theta\cos\phi+\Delta^2\right)^{\frac{1}{2}}$ of the $\ket{\pm}$ spin states as: (a) a function of the azimuthal $0\leq\phi\leq2\pi$ and  polar $0\leq\theta \leq\pi$ field angles for $\Delta = 1$;  and (b) as a function of $\phi$ for $\theta = \frac{\pi}{2}$ and different values of $\Delta$ as indicated. Coordinate axes are provided for reference to figure \ref{fig:centre} where the field angles $\phi=2\phi_{{\cal B}}+\phi_{{\cal E}}$, $\phi_{{\cal E}}$, $\phi_{{\cal B}}$ and $\theta$ are defined. }
\label{fig:solenergies}
\end{center}
\end{figure*}

Figure \ref{fig:solenergies} contains polar plots of the dimensionless splitting parameter $\left(1-2\Delta\sin\theta\cos\phi+\Delta^2\right)^{\frac{1}{2}}$ of the $\ket{\pm}$ spin states as a function of the azimuthal $\phi$ and polar $\theta$ field angles. Clearly the most interesting field configurations have the parameters $\Delta \sim 1$ and $\theta \sim \frac{\pi}{2}$, since for these parameters the energies depend sensitively on the field angles. Such a field configuration was used in the recent electric field sensing demonstration \cite{efield} to sensitively detect both the magnitude and orientation of an applied electric field. Future implementations of the ground state spin as a field sensor should seek to exploit these high sensitivity field configurations.

The matrix representation of the interaction of the ground state spin with an oscillating microwave magnetic field $\vec{M}$ in the basis $\{\ket{0},\ket{-},\ket{+}\}$ is
\begin{widetext}
\begin{eqnarray}
\frac{1}{\hbar}\vec{S}\cdot\vec{{\cal M}} =
\left(\begin{array}{ccc}
0 & \frac{{\cal M}_\perp}{\sqrt{2}}\left(e^{i\frac{\phi_m}{2}}s_{\frac{\theta}{2}}+e^{-i\frac{\phi_m}{2}}c_{\frac{\theta}{2}}\right) & \frac{{\cal M}_\perp}{\sqrt{2}}\left(e^{i\frac{\phi_m}{2}}c_{\frac{\theta}{2}}-e^{-i\frac{\phi_m}{2}}s_{\frac{\theta}{2}}\right) \\
\frac{{\cal M}_\perp}{\sqrt{2}}\left(e^{-i\frac{\phi_m}{2}}s_{\frac{\theta}{2}}+e^{i\frac{\phi_m}{2}}c_{\frac{\theta}{2}}\right) & -{\cal M}_z\cos\theta & {\cal M}_z\sin\theta \\
\frac{{\cal M}_\perp}{\sqrt{2}}\left(e^{-i\frac{\phi_m}{2}}c_{\frac{\theta}{2}}-e^{i\frac{\phi_m}{2}}s_{\frac{\theta}{2}}\right) & {\cal M}_z\sin\theta & {\cal M}_z\cos\theta \\
\end{array}\right) \nonumber\label{eq:mwmatrixrep}
\end{eqnarray}
\end{widetext}
where ${\cal M}_\perp = \sqrt{{\cal M}_x^2+{\cal M}_y^2}$, $\phi_m = 2\phi_{\cal M}+\phi_{\cal E}$ and $\tan\phi_{\cal M} = {\cal M}_y/{\cal M}_x$. The off-diagonals of the above matrix representation indicate that the transitions between the field spin states induced by the oscillating microwave field are also dependent on the static field angles.

\begin{figure}[hbtp]
\begin{center}
\includegraphics[width=1.0\columnwidth] {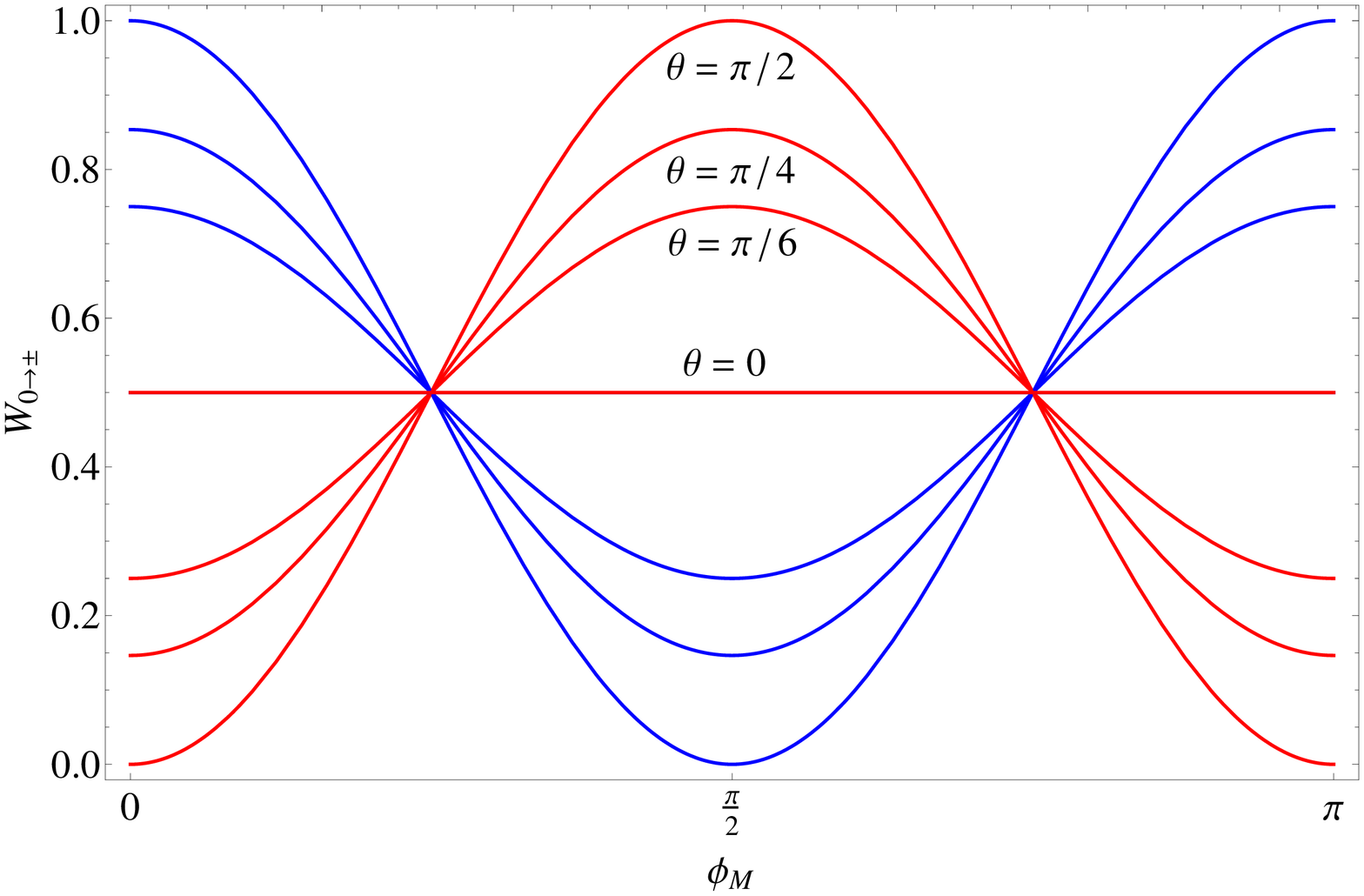}
\caption{(color online) Plots of the normalized transition rates $W_{0\rightarrow-}\propto\frac{1}{2}(1+\sin\theta\cos\phi_m)$ (blue) and $W_{0\rightarrow+}\propto\frac{1}{2}(1-\sin\theta\cos\phi_m)$ (red) between the $\ket{0}$ and $\ket{-}$ and the $\ket{0}$ and $\ket{+}$ spin states, respectively, as functions of the non-axial microwave polarization $\phi_{{\cal M}}$,  given $\phi_{{\cal E}} = 0$ and different values of the static field angle $\theta$. Note that the transition rates are equal for all values of $\phi_{{\cal M}}$ when $\theta = 0$.}
\label{fig:microwavepolarization}
\end{center}
\end{figure}

Using Fermi's golden rule, \cite{MQM} the transition rates $W_{0\rightarrow \pm}$  between the $\ket{0}$ and $\ket{\pm}$ spin states correct to zero-order in the static non-axial magnetic field ${\cal B}_\perp$ are proportional to the absolute square of the off-diagonal elements, such that
\begin{eqnarray}
W_{0\rightarrow \pm} \propto \frac{1}{2}{\cal M}_\perp^2(1\mp\sin\theta\cos\phi_m)
\end{eqnarray}
As the transition rates to the $\ket{\pm}$ spin states depend differently on the static fields for a given microwave polarization, the transitions to the $\ket{\pm}$ spin states can be controlled via the static fields, or conversely for a given static field configuration, the transitions can be controlled via the microwave polarization. Figure \ref{fig:microwavepolarization} depicts the dependence of the transition rates on the microwave polarization and static field configuration. As can be seen, the individual transitions can be selectively excited using orthogonal non-axial microwave polarizations $\phi_{{\cal M}} = \frac{\phi_{{\cal E}}}{2},\frac{\phi_{{\cal E}}}{2}+\frac{\pi}{2}$ when $\theta \sim \frac{\pi}{2}$. Hence, linearly polarized microwaves in conjunction with static field control or (as has been previously demonstrated)\cite{cpolarized} circularly polarized microwaves can be used to selectively excite individual spin transition in situations where the splitting of the $\ket{\pm}$ spin states $(=2{\cal R})$ is too small to selectively excite the individual transitions using microwave frequency selection alone.

\section{Time-evolution of the spin: FID simulation}

\begin{figure}[hbtp]
\begin{center}
\includegraphics[width=0.9\columnwidth] {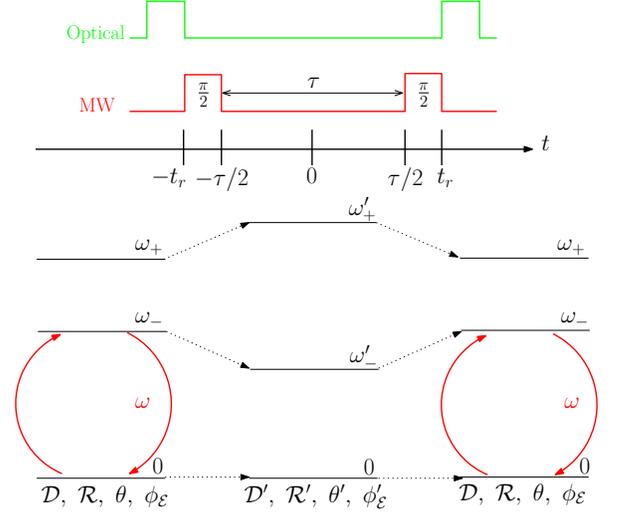}
\caption{(color online) The free induction decay (FID) sequence of optical pulses that polarize the spin at $t=-t_r$ and readout the spin at $t=t_r$, as well as microwave $\pi/2$ pulses that coherently manipulate the spin before and after the period of free evolution $\tau$. The static fields and spin state energies differ during the period of free evolution $({\cal D}^\prime, \ {\cal R}^\prime, \ \theta^\prime, \ \phi_{{\cal E}}^\prime, \ \hbar\omega_\pm^\prime)$ compared to before and after the period of free evolution $({\cal D}, \ {\cal R}, \ \theta, \ \phi_{{\cal E}}, \ \hbar\omega_\pm)$. The microwave field is assumed to selectively excite the transitions between the $\ket{0}$ and $\ket{-}$ spin states with frequency $\omega$. }
\label{fig:FIDsequence}
\end{center}
\end{figure}

Assuming the weak static non-axial magnetic field limit (${\cal B}_\perp \ll {\cal D}$ and ${\cal B}_\perp^2/{\cal D} \ll {\cal R}$), which typically occurs in most of the current applications of the ground state spin, the spin solutions can be used to accurately model the time-evolution of the spin in a simple FID ODMR experiment. As depicted in Fig. \ref{fig:FIDsequence}, the FID sequence is comprised of optical pulses that polarize the spin at $t=-t_r$ and readout the spin at $t=t_r$, as well as microwave $\pi/2$ pulses that coherently manipulate the spin before and after the period of free evolution $\tau$. Note that the MW pulse sequence depicted in Fig. \ref{fig:FIDsequence} differs from the conventional ESR sequence by the final $\pi/2$ pulse, which projects the accumulated phase into a population difference between the $m_s =0$ and $m_s = \pm1$ spin sub-levels.

In the model of the FID experiment, the static fields will be considered to differ during the period of free evolution, such that before and after the period of free evolution the static fields are described by the parameters $({\cal D}, \ {\cal R}, \ \theta, \ \phi_{{\cal E}})$ and the $\ket{\pm}$ spin states have energies $\hbar\omega_\pm$, whilst during the period of free evolution the static fields are described by the parameters $({\cal D}^\prime, \ {\cal R}^\prime, \ \theta^\prime, \ \phi_{{\cal E}}^\prime)$ and the $\ket{\pm}$ spin states have energies $\hbar\omega_\pm^\prime$. For simplicity, the changes in the static field configurations are assumed to be adiabatic and infinitely sharp at $t = \pm\tau/2$ and the microwave field is assumed to selectively excite the transitions between the $\ket{0}$ and $\ket{-}$ spin states with a tuned microwave frequency $\omega\approx \omega_-$.

This model FID experiment is a generalization of the FID experiments that were conducted in the spin's recent electric field sensing demonstration\cite{efield} and one of the spin's magnetic field sensing demonstrations\cite{coherence} and will consequently describe how the combined effects of electric, magnetic and strain fields will influence such sensing demonstrations. However, note that the objective of presenting this model FID experiment in this work is not to discuss sensing techniques, but to provide the necessary theoretical details to discuss the effects of inhomogeneous fields and lattice interactions on the relaxation and dephasing of the spin in the following sections.

The state of the spin  at a given time during the FID sequence $\ket{t}$ can be written in terms of the spin eigenstates of the static fields at that time
\begin{eqnarray}
&&\ket{t}= \nonumber \\
&&\left\{\begin{array}{ll}
c_0(t)\ket{0}+c_-(t)\ket{-}+c_+(t)\ket{+} & -t_r \leq t \leq -\frac{\tau}{2}, \\ & \frac{\tau}{2} \leq t \leq t_r \\
c_0^\prime(t)\ket{0}^\prime+c_-^\prime(t)\ket{-}^\prime+c_+^\prime(t)\ket{+}^\prime &  -\frac{\tau}{2}< t < \frac{\tau}{2} \\ \end{array}\right. \nonumber \\
\end{eqnarray}
where the coefficients $c_i(t)$ and $c_i^\prime(t)$ ($i=0,-,+$) are related at $t=\pm\tau/2$ by the basis transformation ${\cal T}: \{\ket{0}, \ket{-},\ket{+}\} \rightarrow \{\ket{0}^\prime, \ket{-}^\prime,\ket{+}^\prime\}$ given by the matrix

\begin{eqnarray}
&&{\cal T} = \notag \\
&&\left(\begin{array}{ccc}
1 & 0 & 0 \\
0 & c_{\delta\phi_{{\cal E}}}c_{\delta\theta_-}-is_{\delta\phi_{{\cal E}}}c_{\delta\theta_+} & -c_{\delta\phi_{{\cal E}}}s_{\delta\theta_-}+is_{\delta\phi_{{\cal E}}}s_{\delta\theta_+} \\
0 & c_{\delta\phi_{{\cal E}}}s_{\delta\theta_-}+is_{\delta\phi_{{\cal E}}}s_{\delta\theta_+} & c_{\delta\phi_{{\cal E}}}c_{\delta\theta_-}+is_{\delta\phi_{{\cal E}}}c_{\delta\theta_+} \\
\end{array}\right)\notag \\
\end{eqnarray}
where $\delta\phi_{{\cal E}} = \frac{1}{2}(\phi_{{\cal E}}^\prime-\phi_{{\cal E}})$ and $\delta\theta_\pm = \frac{1}{2}(\theta^\prime\pm\theta)$.

Introducing the density operator $\hat{\rho}(t) = \ketbra{t}{t}$, the matrix representation of $\hat{\rho}(t)$ in the spin eigenstate basis of the fields at $t$ $(\{\ket{0}, \ket{-},\ket{+}\}$ or $\{\ket{0}^\prime, \ket{-}^\prime,\ket{+}^\prime\}$) is
\begin{eqnarray}
\hat{\rho}(t) = \left\{\begin{array}{ll}
\rho(t) & -t_r \leq t \leq -\frac{\tau}{2}, \  \frac{\tau}{2} \leq t \leq t_r \\
\rho^\prime(t) &  -\frac{\tau}{2}< t < \frac{\tau}{2} \\ \end{array}\right.
\end{eqnarray}
where the elements of the matrices $\rho(t)$ and $\rho^\prime(t)$ are $\rho_{ij}(t)=c_i(t)c_j^\ast(t)$ and $\rho_{ij}^\prime(t)=c_i^\prime(t)c_j^{\prime\ast}(t)$ respectively, and at $t = \pm\frac{\tau}{2}$ the matrices are related by the transformations $\rho^\prime(-\frac{\tau}{2})={\cal T}^{-1}\rho(-\frac{\tau}{2}){\cal T}$ and $\rho(\frac{\tau}{2})={\cal T}\rho^\prime(\frac{\tau}{2}){\cal T}^{-1}$.

In order to simplify the treatment of the interaction of the spin with the microwave pulses, the rotating wave approximation can be adopted and the density operator transformed into the rotating reference frame, such that $\rho(t) \rightarrow \tilde{\rho}(t)$ and $\rho^\prime(t) \rightarrow \tilde{\rho}^\prime(t)$, where $\tilde{\rho}_{\pm0}^\ast(t)=\tilde{\rho}_{0\pm}(t) = \rho_{0\pm}(t)e^{-i\omega t}$, $\tilde{\rho}_{\pm0}^{\prime\ast}(t)=\tilde{\rho}_{0\pm}^\prime(t) = \rho_{0\pm}^\prime(t)e^{-i\omega t}$, and $\tilde{\rho}_{ij}(t) = \rho_{ij}(t)$ and $\tilde{\rho}_{ij}^\prime(t) = \rho_{ij}^\prime(t)$ for the other density matrix elements. \cite{louden} The effect of the $\pi/2$ microwave pulses can therefore be described by $\tilde{\rho}(-\frac{\tau}{2}) = e^{-iJ\frac{\pi}{2}}\tilde{\rho}(-t_r)e^{iJ\frac{\pi}{2}}$ and $\tilde{\rho}(t_r) = e^{-iJ\frac{\pi}{2}}\tilde{\rho}(\frac{\tau}{2})e^{iJ\frac{\pi}{2}}$, where using the microwave matrix representation (\ref{eq:mwmatrixrep})
\begin{eqnarray}
J = \frac{1}{2}\left(\begin{array}{ccc}
0 & e^{i\Omega} & 0 \\
e^{-i\Omega} & 0 & 0 \\
0 & 0 & \frac{4}{\pi}\delta\omega_{+-}\delta t_{\frac{\pi}{2}} \\
\end{array}\right)
\end{eqnarray}
$\tan\Omega = \frac{\sin\left(\frac{\theta}{2}-\frac{\pi}{4}\right)}{\sin\left(\frac{\theta}{2}+\frac{\pi}{4}\right)}\tan\frac{\phi_m}{2}$, $\delta\omega_{+-} = \omega_+-\omega_-$, and $\delta t_{\frac{\pi}{2}}$ is the duration of the $\pi/2$ pulse.

Due to the process of optical spin-polarization, the first optical pulse of the FID sequence will incoherently polarize the ground state spin such that at $t=-t_r$,
\begin{eqnarray}
\tilde{\rho}(-t_r) = \left(\begin{array}{ccc} p_0 & 0 & 0 \\ 0 & p_- & 0 \\ 0 & 0 & p_+ \\ \end{array}\right)
\end{eqnarray}
where $p_i$ are the populations of the spin sub-levels and $p_0+p_-+p_+=1$. Note that since it is believed that the spin-polarization process does not discriminate between the population of the $m_s = \pm1$ sub-levels, it is expected that $p_+ \approx p_-$. The effect of the first $\pi/2$ microwave pulse is
\begin{eqnarray}
&&\tilde{\rho}(-\frac{\tau}{2}) = e^{-iJ\frac{\pi}{2}}\tilde{\rho}(-t_r)e^{iJ\frac{\pi}{2}} = \nonumber \\
&&\left(\begin{array}{ccc}
p_0-\frac{1}{2}\delta p & \frac{i}{2}e^{i\Omega}\delta p & 0 \\
\frac{-i}{2}e^{-i\Omega}\delta p & p_0-\frac{1}{2}\delta p & 0 \\
 0 & 0 & p_-
\end{array}\right)
\end{eqnarray}
where $\delta p = p_0-p_-$ and the simplifying assumption $p_+ \approx p_-$ has been made. After the sudden change in field configuration at $t=-\frac{\tau}{2}$, $\tilde{\rho}(-\frac{\tau}{2})$ is transformed into
\begin{eqnarray}
&&\tilde{\rho}^\prime(-\frac{\tau}{2}) = {\cal T}^{-1}\tilde{\rho}(-\frac{\tau}{2}){\cal T}= \nonumber \\
&&\left(\begin{array}{ccc}
p_0-\frac{1}{2}\delta p & \frac{i}{2}e^{i\Omega}{\cal T}_{--}\delta p & \frac{i}{2}e^{i\Omega}{\cal T}_{-+}\delta p \\
\frac{-i}{2}e^{-i\Omega}{\cal T}_{--}^\ast\delta p & p_-+\frac{1}{2}|{\cal T}_{--}|^2\delta p & \frac{1}{2}{\cal T}_{--}^\ast{\cal T}_{-+}\delta p\\
\frac{-i}{2}e^{-i\Omega}{\cal T}_{-+}^\ast\delta p & \frac{1}{2}{\cal T}_{--}{\cal T}_{-+}^\ast\delta p & p_-+\frac{1}{2}|{\cal T}_{-+}|^2\delta p \\
\end{array}\right) \nonumber \\
\end{eqnarray}
where ${\cal T}_{--} = \cos\delta\phi_{{\cal E}}\cos\delta\theta_--i\sin\delta\phi_{{\cal E}}\cos\delta\theta_+$ and ${\cal T}_{-+} = -\cos\delta\phi_{{\cal E}}\sin\delta\theta_-+i\sin\delta\phi_{{\cal E}}\sin\delta\theta_+$ are elements of the basis transformation matrix.

The coherent time evolution of the ground state spin during the free evolution period is governed by the Schr$\mathrm{\ddot{o}}$dinger equation $\frac{d}{dt}\hat{\rho}(t) = \frac{i}{\hbar}\left[\hat{\rho}(t),\hat{H}_{gs}^\prime\right]$. \cite{louden} However, the spin also interacts with incoherent time-dependent perturbations such as crystal vibrations, the thermal radiation field and fluctuating fields from local magnetic and electric impurities during the free evolution period. These incoherent perturbations induce transitions between the spin states that lead to spin relaxation $\gamma_{ij}^r$ and dephasing rates $\gamma_{ij}^p$, which are characterized by the $T_1$ and $T_2$ times of the spin, respectively.\cite{EPR} Accounting for both the coherent and incoherent evolution of the spin, \cite{louden} the following matrix equation is obtained
\begin{widetext}
\begin{eqnarray}
\frac{d}{dt}\tilde{\rho}^\prime(t) =
\left(\begin{array}{ccc}
-(\gamma_{0-}^r+\gamma_{0+}^r)\tilde{\rho}_{00}^\prime& (i\delta\omega_-^\prime-\gamma_{-0}^p)\tilde{\rho}_{0-}^\prime & (i\delta\omega_{+}^\prime-\gamma_{+0}^p)\tilde{\rho}_{0+}^\prime\\
+\gamma_{-0}^r\tilde{\rho}_{--}^\prime+\gamma_{+0}^r\tilde{\rho}_{++}^\prime & & \\
(-i\delta\omega_-^\prime-\gamma_{-0}^p)\tilde{\rho}_{-0}^\prime & -(\gamma_{-0}^r+\gamma_{-+}^r)\tilde{\rho}_{--}^\prime & (i\delta\omega_{+-}^\prime-\gamma_{+-}^p)\tilde{\rho}_{-+}^\prime \\
 & +\gamma_{0-}^r\tilde{\rho}_{00}^\prime+\gamma_{+-}^r\tilde{\rho}_{++}^\prime & \\
(-i\delta\omega_{+}^\prime-\gamma_{+0}^p)\tilde{\rho}_{+0}^\prime & (i\delta\omega_{+-}^\prime-\gamma_{+-}^p)\tilde{\rho}_{+-}^\prime & -(\gamma_{+0}^r+\gamma_{+-}^r)\tilde{\rho}_{++}^\prime \\
 & & +\gamma_{0+}^r\tilde{\rho}_{00}^\prime+\gamma_{-+}^r\tilde{\rho}_{--}^\prime \\
\end{array}\right)
\end{eqnarray}
\end{widetext}
where $\delta\omega_\pm^\prime=\omega_\pm^\prime-\omega$ and $\delta\omega_{+-}^\prime=\omega_+^\prime-\omega_-^\prime$. Note that the coherent coupling between the NV spin and proximal impurity spins will not be considered in this work.

The solutions of the off-diagonal elements at $t=\frac{\tau}{2}$ are simply
\begin{eqnarray}
\tilde{\rho}_{-0}^{\prime\ast}(\frac{\tau}{2}) = \tilde{\rho}_{0-}^\prime(\frac{\tau}{2}) = \tilde{\rho}_{0-}^\prime(-\frac{\tau}{2})e^{-\gamma_{-0}^p\tau}e^{i\delta\omega_-^\prime\tau} \nonumber \\
\tilde{\rho}_{+0}^{\prime\ast}(\frac{\tau}{2}) = \tilde{\rho}_{0+}^\prime(\frac{\tau}{2}) = \tilde{\rho}_{0+}^\prime(-\frac{\tau}{2})e^{-\gamma_{+0}^p\tau}e^{i\delta\omega_{+}^\prime\tau} \nonumber \\
\tilde{\rho}_{+-}^{\prime\ast}(\frac{\tau}{2}) = \tilde{\rho}_{-+}^\prime(\frac{\tau}{2}) = \tilde{\rho}_{-+}^\prime(-\frac{\tau}{2})e^{-\gamma_{+-}^p\tau}e^{i\delta\omega_{+-}^\prime\tau}
\end{eqnarray}
whilst the solutions of the diagonal elements $\tilde{\rho}_{00}^\prime(\frac{\tau}{2})$, $\tilde{\rho}_{--}^\prime(\frac{\tau}{2})$ and $\tilde{\rho}_{++}^\prime(\frac{\tau}{2})$ are more complicated, but can still be obtained analytically as combinations of exponential functions of the relaxation rates. It follows that after the second sudden change in the static fields at $t = \frac{\tau}{2}$, $\tilde{\rho}(\frac{\tau}{2}) = {\cal T}\tilde{\rho}^\prime(\frac{\tau}{2}){\cal T}^{-1}$, and after the second $\pi/2$ microwave pulse at $t = t_r$, $\tilde{\rho}(t_r) = e^{-iJ\frac{\pi}{2}}\tilde{\rho}(\frac{\tau}{2})e^{iJ\frac{\pi}{2}}$. The second optical pulse at $t=t_r$ reads out the proportion of the population in the $m_s = 0$ sub-level, such that the optical emission intensity $I(\tau) \propto \tilde{\rho}_{00}(t_r)$, where
\begin{eqnarray}
\tilde{\rho}_{00}(t_r) & = & \frac{1}{2}\left[\tilde{\rho}_{00}^\prime(\frac{\tau}{2})+|{\cal T}_{--}|^2\tilde{\rho}_{--}^\prime(\frac{\tau}{2})+|{\cal T}_{-+}|^2\tilde{\rho}_{++}^\prime(\frac{\tau}{2})\right]
\nonumber \\
&& -\frac{\delta p}{2}\left[|{\cal T}_{--}|^2e^{-\gamma_{-0}^p\tau}\cos\delta\omega_-^\prime\tau \right. \nonumber \\
&&\left.+|{\cal T}_{-+}|^2e^{-\gamma_{+0}^p\tau}\cos\delta\omega_+^\prime\tau\right. \nonumber \\
&&\left.-|{\cal T}_{--}|^2|{\cal T}_{-+}|^2e^{-\gamma_{+-}^p\tau}\cos\delta\omega_{+-}^\prime\tau\right] \nonumber \\
& = & \frac{1}{2}N(\tau)-\frac{\delta p}{2}O(\tau)
\end{eqnarray}

The first term $N(\tau)$ is not oscillatory and depends only on the diagonal elements $\tilde{\rho}_{00}^\prime(-\frac{\tau}{2})$, $\tilde{\rho}_{--}^\prime(-\frac{\tau}{2})$ and $\tilde{\rho}_{++}^\prime(-\frac{\tau}{2})$ and the relaxation rates $\gamma_{ij}^r$. The second term $O(\tau)$ has oscillatory components with frequencies corresponding to the different frequency shifts ($\delta\omega_\pm^\prime$ and $\delta\omega_{+-}^\prime$) and the contributions of each oscillatory component are dependent on the state couplings ($|{\cal T}_{--}|$ and $|{\cal T}_{-+}|$) and the dephasing rates $\gamma_{ij}^p$. The oscillatory term therefore offers a great deal of information about the spin eigenstates and their energies during the period of free evolution. Furthermore, since the observed $T_1$ and $T_2$ times of the ground state spin typically differ by at least on order of magnitude,\cite{coherence} the change in the non-oscillatory term over the lifetime of the oscillatory term is negligible, and thus can be effectively ignored in an observation of the oscillatory term.

The oscillatory term can be observed by conducting ODMR measurements of an ensemble of spins or conducting many ODMR measurements of a single spin. For a given measurement of an ensemble of spins, the field parameters $({\cal D},{\cal R}, \theta, \phi_{{\cal E}})$ and $({\cal D}^\prime,{\cal R}^\prime, \theta^\prime, \phi_{{\cal E}}^\prime)$ will potentially differ for each spin within the ensemble due to inhomogeneities in the fields. Likewise, for an ensemble of measurements of a single spin, the field parameters will potentially differ between each measurement due to differences in the preparation of the spin and the fields. These ensemble inhomogeneities introduce an additional dephasing decay in the observation of the oscillatory term and can be accounted for by introducing statistical distribution functions of the field parameters and calculating the expectation value $\langle O(\tau) \rangle$. \cite{pulsedEPR}

\section{The effects of inhomogeneous fields}

The total dephasing rate of the spin due to interactions with incoherent time-dependent fields and inhomogeneous static fields is characterized by the $T_2^\ast$ time of the spin.\cite{EPR} In the recent electric field sensing demonstration, \cite{efield} it was observed that the $T_2^\ast$ time of the ground state spin was highly dependent on the field configuration, such that it obtained a maximum in the absence of an axial magnetic field and sharply decreased as the axial magnetic field was increased at a rate that was inversely related to the non-axial electric-strain field. This observation highlighted the significant influence that the static fields have on the dephasing of the spin and the potential to control the susceptability of the spin to different noise sources. The dephasing due to static field inhomogenieites will be discussed in this section and the dependence of the incoherent dephasing rates $\gamma_{ij}^p$ on the static fields will be discussed in the next section.

Since the ground state spin interacts very weakly with axial electric-strain fields, the effect of the variation of ${\cal D}$ and ${\cal D}^\prime$ between the measurements of a single spin will be negligible compared to the variations in the other field parameters. Note that this is not necessarily the case for an ensemble of spins because local strain fields can vary significantly between lattice sites. Nor is it necessarily the case for single spins or ensembles of spins if temperature varies appreciably during the conduct of the measurements.\cite{dtempdep} Nevertheless, the variations in ${\cal D}$ and ${\cal D}^\prime$ are considered negligible in the following. It is reasonable to expect that the field parameters ${\cal R}$ and ${\cal R}^\prime$ will have statistical distributions $f({\cal R}/\hbar,\mu_{{\cal R}},\sigma_{{\cal R}})$ and $f({\cal R}^\prime/\hbar,\mu_{{\cal R}^\prime},\sigma_{{\cal R}^\prime})$ that have the same distribution function $f$, but different mean values $\mu_{{\cal R}}$ and $\mu_{{\cal R}^\prime}$ and different variances $\sigma_{{\cal R}}$ and $\sigma_{{\cal R}^\prime}$ in units of frequency.

The distribution functions of the frequency shifts $\delta\omega_\pm^\prime$ and $\delta\omega_{+-}^\prime$ can be constructed using the distributions of ${\cal R}$ and ${\cal R}^\prime$ as\cite{statistics}
\begin{eqnarray}
&&F_-(\delta\omega_-^\prime, \mu_-,\sigma_-)  =  \nonumber \\
&& \int_0^\infty f({\cal R}/\hbar,\mu_{{\cal R}},\sigma_{{\cal R}})f(\delta\omega_-^\prime+{\cal R}/\hbar,\mu_{{\cal R}^\prime},\sigma_{{\cal R}^\prime})d{\cal R}/\hbar \nonumber \\
&&F_+(\delta\omega_+^\prime, \mu_+,\sigma_+)  =  \nonumber \\
&&\int_0^\infty f({\cal R}/\hbar,\mu_{{\cal R}},\sigma_{{\cal R}})f(\delta\omega_+^\prime-{\cal R}/\hbar,\mu_{{\cal R}^\prime},\sigma_{{\cal R}^\prime})d{\cal R}/\hbar \nonumber \\
&&F_{+-}(\delta\omega_{+-}^\prime,\mu_{+-},\sigma_{+-})  = \frac{1}{2}f(\frac{1}{2}\delta\omega_{+-}^\prime,\mu_{{\cal R}^\prime},\sigma_{{\cal R}^\prime}) \label{eq:Fdistribution}
\end{eqnarray}
where the explicit expressions for the means $\mu_\pm$ and $\mu_{+-}$ and the variances $\sigma_\pm$ and $\sigma_{+-}$ in terms of $\mu_{{\cal R}}$, $\mu_{{\cal R}^\prime}$, $\sigma_{{\cal R}}$ and $\sigma_{{\cal R}^\prime}$ depend on the distribution $f$. For example, if $f$ is the normal distribution, then the expressions are simply $\mu_\pm = \delta{\cal D}/\hbar+\mu_{{\cal R}}\pm\mu_{{\cal R}^\prime}$, $\mu_{+-}=2\mu_{{\cal R}^\prime}$, $\sigma_\pm = \sigma_{{\cal R}}+\sigma_{{\cal R}^\prime}$ and $\sigma_{+-}=2\sigma_{{\cal R}^\prime}$.\cite{statistics}

Using the distributions of the frequency shifts, the expectation value of the FID oscillatory term for an ensemble of measurements of a single spin is
\begin{eqnarray}
&& \langle O(\tau) \rangle \nonumber \\
&& = \langle |{\cal T}_{--}|^2\rangle e^{-\gamma_{-0}^p\tau}\int_{-\infty}^\infty\cos\delta\omega_-^\prime\tau F_-(\delta\omega_-^\prime,\mu_-,\sigma_-)d\delta\omega_{-}^\prime \nonumber \\
&&+ \langle|{\cal T}_{-+}|^2\rangle e^{-\gamma_{+0}^p\tau}\int_{-\infty}^\infty\cos\delta\omega_+^\prime\tau F_+(\delta\omega_+^\prime,\mu_+,\sigma_+)d\delta\omega_{+}^\prime \nonumber \\
&&- \langle|{\cal T}_{--}|^2|{\cal T}_{-+}|^2\rangle e^{-\gamma_{+-}^p\tau}\int_{-\infty}^\infty\frac{1}{2}\cos\delta\omega_{+-}^\prime\tau  \nonumber \\
&& \ \ \ \ \ \ \ \ \ \ \ \ \  \ \ \ \ \ \ \ \ \ \ \ F_{+-}(\frac{1}{2}\delta\omega_{+-}^\prime,\mu_{+-},\sigma_{+-})d\delta\omega_{+-}^\prime
\end{eqnarray}
Note that the expectation values of the state couplings involve just the field angles $\delta\phi_{{\cal E}}$ and $\delta\theta_\pm$.

The above expression demonstrates that $\langle O(\tau)\rangle$ is potentially complicated for general state couplings and distribution functions and that it is difficult to extract all of the information encoded in the oscillatory term. A clearer analysis of the oscillatory term is obtained by performing a Fourier cosine transformation,
\begin{eqnarray}
&&\langle O(\nu) \rangle \nonumber \\
&& = \frac{2}{\pi}\int_0^\infty\langle O(\tau)\rangle \cos\nu\tau d\tau\nonumber \\
&& =  \langle |{\cal T}_{--}|^2\rangle \int_{-\infty}^\infty L(\nu,\delta\omega_-^\prime,\gamma_{-0}^p) F_-(\delta\omega_-^\prime,\mu_-,\sigma_-)d\delta\omega_{-}^\prime \nonumber \\
&&+ \langle|{\cal T}_{-+}|^2\rangle\int_{-\infty}^\infty L(\nu,\delta\omega_+^\prime,\gamma_{+0}^p) F_+(\delta\omega_+^\prime,\mu_+,\sigma_+)d\delta\omega_{+}^\prime \nonumber \\
&&- \langle|{\cal T}_{--}|^2|{\cal T}_{-+}|^2\rangle \int_{-\infty}^\infty\frac{1}{2}L(\nu,\delta\omega_{+-}^\prime,\gamma_{+-}^p) \nonumber \\
&& \ \ \ \ \ \ \ \ \ \ \ \ \  \ \ \ \ \ \ \ \ \ \ \ F_{+-}(\frac{1}{2}\delta\omega_{+-}^\prime,\mu_{+-},\sigma_{+-})d\delta\omega_{+-}^\prime  \label{eq:fourierspectrum}
\end{eqnarray}
where $L(\nu,x,\gamma) = \frac{1}{\pi}\left(\frac{\gamma}{\gamma^2+(\nu-x)^2}+\frac{\gamma}{\gamma^2+(\nu+x)^2}\right)$
is a sum of Lorentzian distributions. Consequently, it is clear that $\langle O(\nu) \rangle$ is comprised of a collection of lines at $\nu=$$\pm\mu_-$, $\pm\mu_+$, $\pm\mu_{+-}$ which have composite lineshapes of $L$ and the distribution functions of frequency shifts. The Fourier analysis of the oscillatory term therefore provides the frequency shifts, the state couplings as well as information about the distributions of the field parameters from the locations, intensities and shapes of the lines. This additional information encoded in the lineshapes can be used to infer details about the statistics of the local environment of the spin, a notion which (through a different approach) forms the basis of the recent proposals of decoherence imaging. \cite{decoherence1,decoherence2,decoherence3,decoherence4,decoherence5}

The distribution function $f$ of the field parameters ${\cal R}$ and ${\cal R}^\prime$ can be itself constructed from the distributions of the electric-strain ${\cal E}_\perp$ and magnetic ${\cal B}_z$ field components. Let the distributions of the electric-strain and magnetic field components be $\epsilon$ and $\beta$ respectively, then\cite{statistics}
\begin{eqnarray}
&&f({\cal R}/\hbar,\mu_{{\cal R}},\sigma_{{\cal R}}) = \nonumber \\
&&\frac{d}{d{\cal R}}\int_{{\cal R}\geq \sqrt{{\cal B}_z^2+{\cal E}_\perp^2}}\epsilon(\frac{{\cal E}_\perp}{\hbar},\mu_{{\cal E}},\sigma_{{\cal E}})\beta(\frac{{\cal B}_z}{\hbar},\mu_{{\cal B}},\sigma_{{\cal B}})d\frac{{\cal E}_\perp}{\hbar} d\frac{{\cal B}_z}{\hbar} \nonumber \\
&&f({\cal R}^\prime/\hbar,\mu_{{\cal R}^\prime},\sigma_{{\cal R}^\prime}) = \nonumber \\
&&\frac{d}{d{\cal R}^\prime}\int_{{\cal R}^\prime\geq \sqrt{{\cal B}_z^2+{\cal E}_\perp^2}}\epsilon(\frac{{\cal E}_\perp}{\hbar},\mu_{{\cal E}}^\prime,\sigma_{{\cal E}})\beta(\frac{{\cal B}_z}{\hbar},\mu_{{\cal B}}^\prime,\sigma_{{\cal B}})d\frac{{\cal E}_\perp}{\hbar} d\frac{{\cal B}_z}{\hbar} \nonumber \\
\end{eqnarray}
where it has been assumed that the variances of the electric-strain and magnetic field components are independent of their mean values.

The above construction can be demonstrated using the simple case where the variance of the non-axial electric-strain field $\sigma_{{\cal E}}$ is negligible compared to the variance of the axial magnetic field $\sigma_{{\cal B}}$ and the mean values of the field components $\mu_{{\cal E}}$ and $\mu_{{\cal B}}$. Due to the dominance of paramagnetic impurities over electric impurities in diamond,\cite{coherence} this simple case is applicable to most applications of the ground state spin. Modeling the electric-strain field distribution $\epsilon = \delta({\cal E}_\perp/\hbar-\mu_{{\cal E}})$ by a delta function and the magnetic field distribution $\beta = {\cal N}({\cal B}_z/\hbar,\mu_{{\cal B}},\sigma_{{\cal B}})$ by a normal distribution  ${\cal N}(x,\mu,\sigma) = e^{-\frac{(x-\mu)^2}{2\sigma^2}}/\sqrt{2\pi\sigma^2}$, the distribution function of ${\cal R}$ is
\begin{eqnarray}
&& f({\cal R}/\hbar,\mu_{{\cal R}},\sigma_{{\cal R}}) =\nonumber \\
&& N_{{\cal R}}(\mu,\sigma)\left\{\begin{array}{ll}
0 & u<1 \\
\frac{\sqrt{u^2-1}}{u}\left[{\cal N}(\sqrt{u^2-1},\mu,\sigma)\right. & \\
 \left.+{\cal N}(-\sqrt{u^2-1},\mu,\sigma)\right] & u\geq1 \\
\end{array}\right.\label{eq:fdistribution}
\end{eqnarray}
where $u = {\cal R}/\hbar\mu_{{\cal E}}$, $\mu = \mu_{{\cal B}}/\mu_{{\cal E}}$, $\sigma = \sigma_{{\cal B}}/\mu_{{\cal E}}$ and $N_{\cal R}(\mu,\sigma)$ is a normalization constant. Note that an analogous expression can be obtained for ${\cal R}^\prime$ by substituting the respective mean values and variances of the field components.

\begin{figure}[hbtp]
\begin{center}
\mbox{
\subfigure[]{\includegraphics[width=0.8\columnwidth] {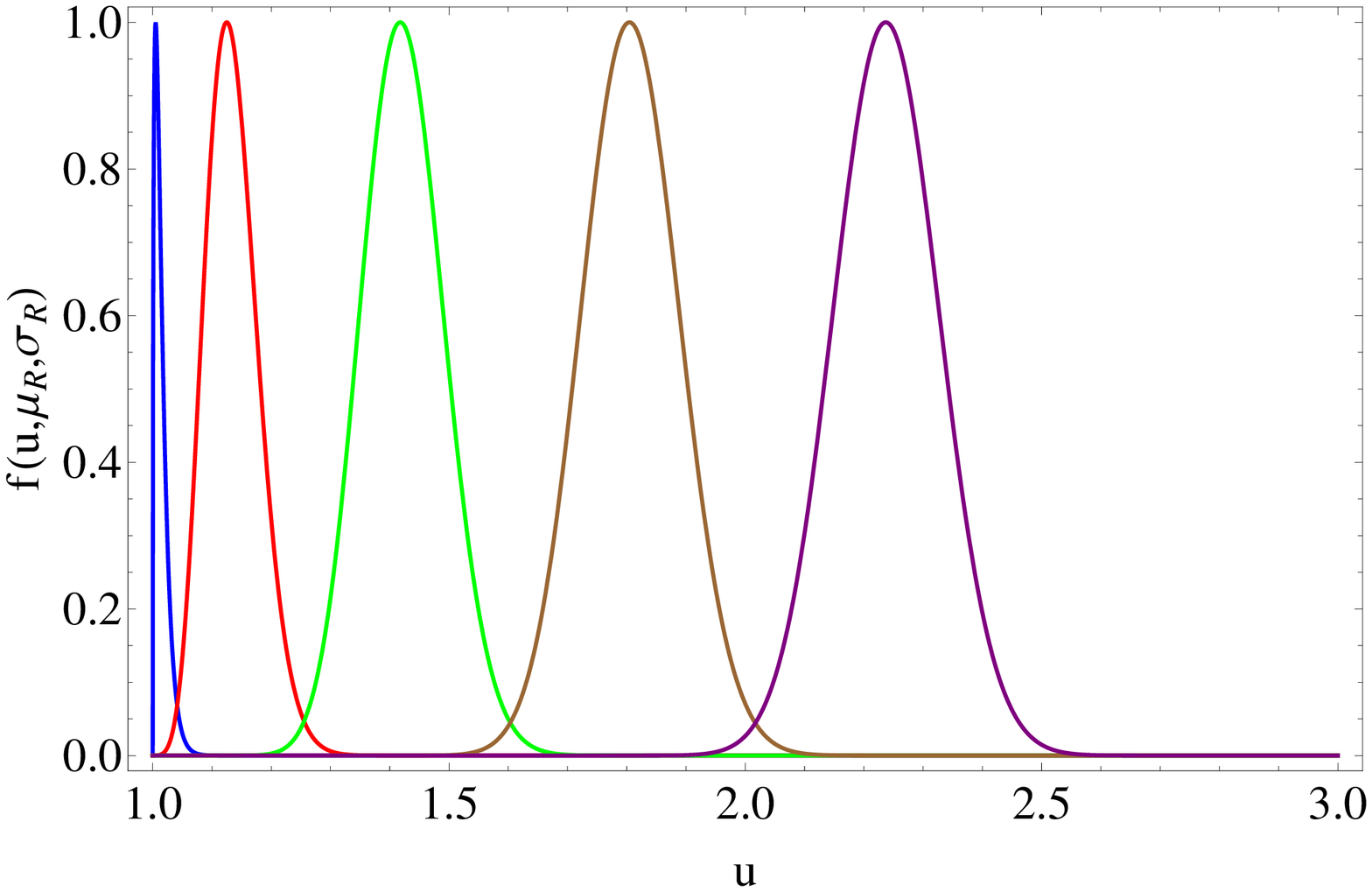}}
}
\mbox{
\subfigure[]{\includegraphics[width=0.8\columnwidth] {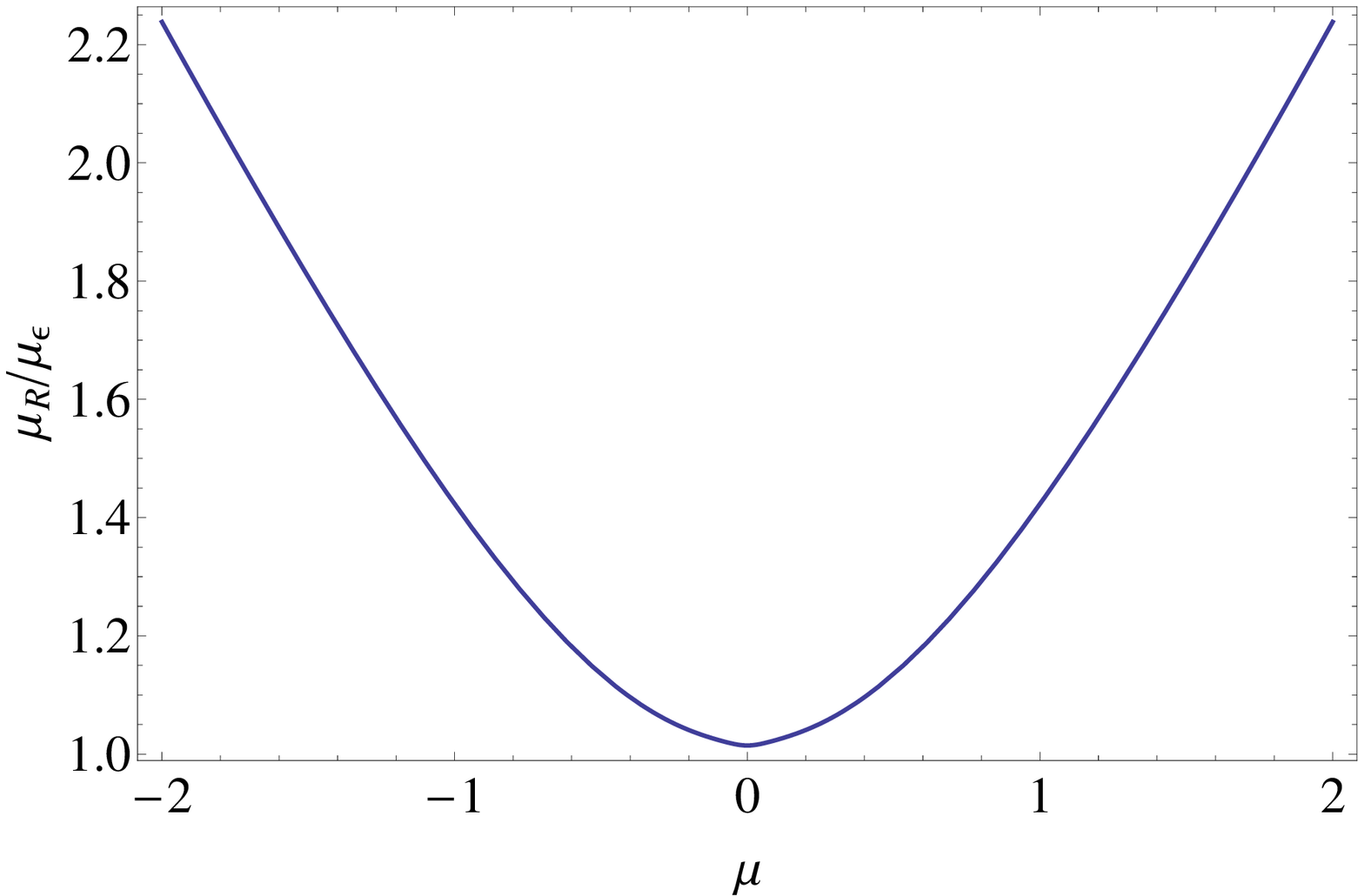}}}
\mbox{
\subfigure[]{\includegraphics[width=0.8\columnwidth] {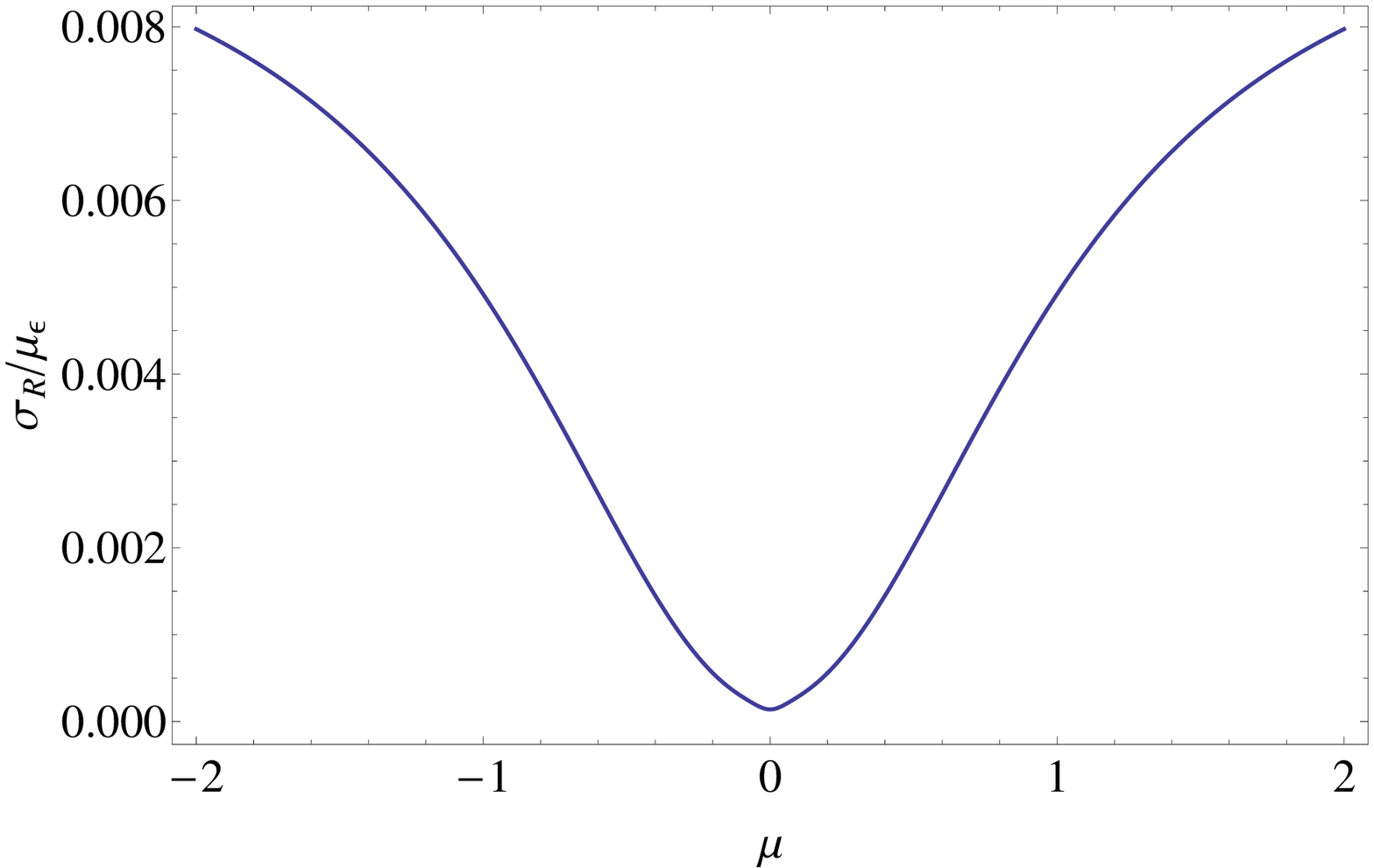}}
}
\caption{(color online) The distribution $f({\cal R}/\hbar,\mu_{{\cal R}},\sigma_{{\cal R}})$ constructed for the simple case where the distribution of the non-axial electric-strain component has negligible variance compared to the variance of the axial magnetic field component $\sigma_{{\cal B}}$ and the mean values of the field components $\mu_{{\cal E}}$ and $\mu_{{\cal B}}$. (a) plots of the distribution as a function of $u = {\cal R}/\hbar\mu_{{\cal E}}$ for $\mu = \mu_{{\cal B}}/\mu_{{\cal E}}= 0,1/2,1,3/2,2$ (in sequential order left to right) and the same variance $\sigma = \sigma_{{\cal B}}/\mu_{{\cal E}} = 0.1$. (b) and (c) are plots of the mean $\mu_{{\cal R}}/\mu_{{\cal E}}$ and variance $\sigma_{{\cal R}}/\mu_{{\cal E}}$ of the distribution as a function of the axial magnetic field mean $\mu$ given $\sigma = 0.1$. }
\label{fig:fdistribution}
\end{center}
\end{figure}

Figure \ref{fig:fdistribution} (a) contains plots of the above distribution of ${\cal R}$ for the relative magnetic field variance $\sigma=0.1$ and for different relative axial magnetic field mean values $\mu$. As demonstrated by  figure \ref{fig:fdistribution} (b) and (c), the mean of the distribution of ${\cal R}$ varies as $\mu_{{\cal R}}/\mu_{{\cal E}} = \sqrt{\mu^2+1}$, which would be expected from the relationship ${\cal R} =\sqrt{{\cal B}_z^2+{\cal E}_\perp^2}$, and that the variance $\sigma_{{\cal R}}/\mu_{{\cal E}}$ depends sensitively on the relative axial magnetic field mean $\mu $, except for $\mu \gg 1$, where the variance becomes approximately independent of $\mu$.

Consider a FID experiment where the static fields do not differ in the period of free evolution. Such FID experiments are typically used to measure the $T_2^\ast$ time of the ground state spin via the width of the single line that occurs at $\delta\omega_-^\prime = 0$  in the Fourier spectrum of the oscillatory term. Noting that for such an experiment $\langle |{\cal T}_{-+}|^2\rangle = \langle |{\cal T}_{--}|^2{\cal T}_{-+}|^2\rangle = 0$,  $\mu_{{\cal R}} = \mu_{{\cal R}^\prime}$, $\sigma_{{\cal R}} = \sigma_{{\cal R}^\prime}$ and $\delta {\cal D} = 0$, the expression for the Fourier spectrum (\ref{eq:fourierspectrum}) simplifies to
\begin{eqnarray}
\langle O(\nu)\rangle = \int_{-\infty}^\infty L(\nu,\delta\omega_-^\prime,\gamma_{-0}^p)F_-(\delta\omega_-^\prime,\mu_-,\sigma_-)d\delta\omega_-^\prime
\end{eqnarray}
where
\begin{eqnarray}
&&F_-(\delta\omega_-^\prime, \mu_-,\sigma_-)  = \nonumber \\
&&\int_0^\infty f({\cal R}/\hbar,\mu_{{\cal R}},\sigma_{{\cal R}})f(\delta\omega_-^\prime+{\cal R}/\hbar,\mu_{{\cal R}},\sigma_{{\cal R}})d{\cal R}/\hbar \nonumber \\
\end{eqnarray}
and assuming the simple case of negligible electric-strain field variance, the distribution $f$ is given by (\ref{eq:fdistribution}). Since the lineshape is a composition of $L$ and $F_-$, the width of the line, and thus the $T_2^\ast$ time of the spin, will depend on both the dephasing rate $\gamma_{-0}^p$ and the electric-strain and magnetic field distribution parameters that form $F_-$.

\begin{figure}[hbtp]
\begin{center}
\mbox{
\subfigure[]{\includegraphics[width=0.95\columnwidth] {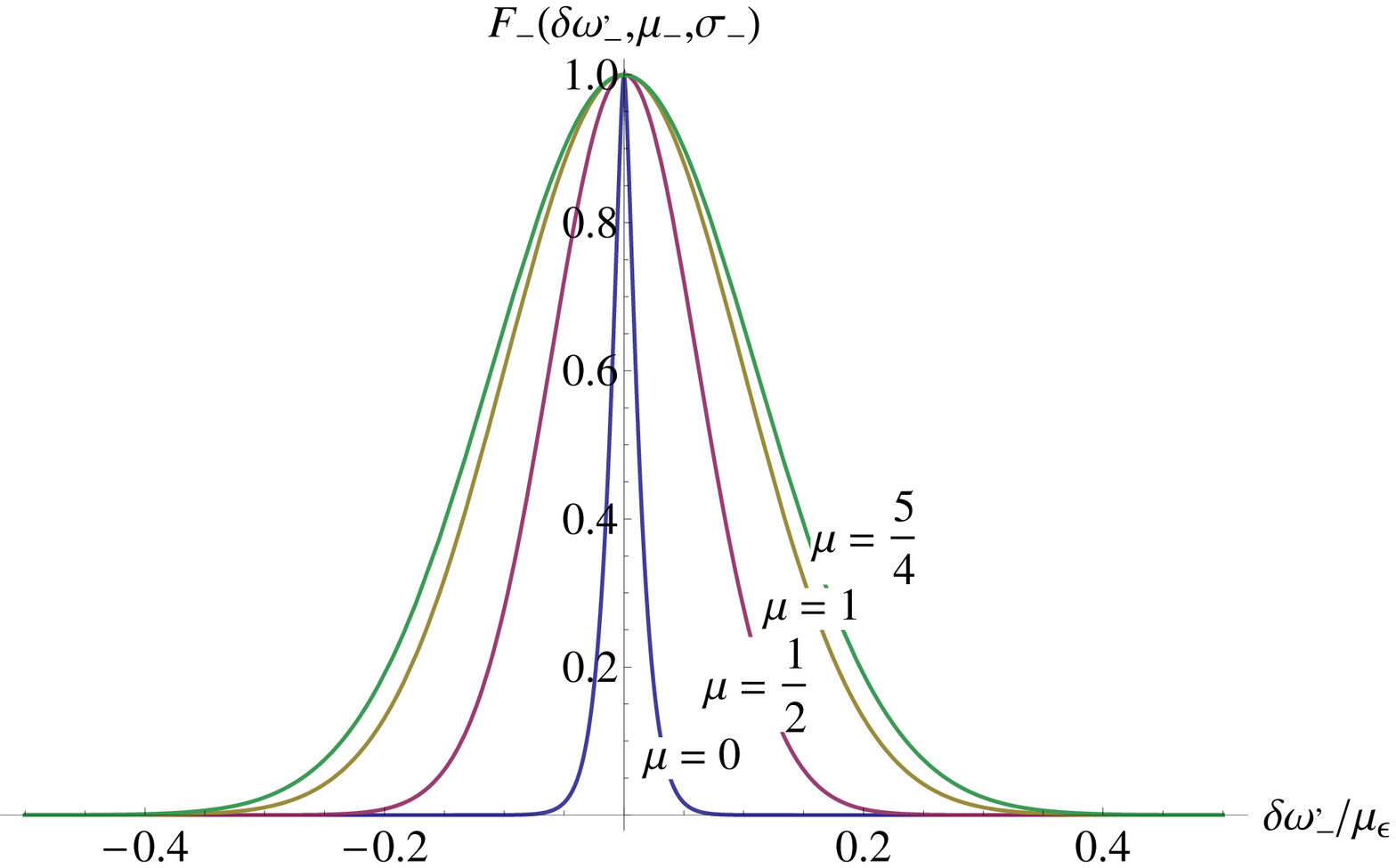}}}
\mbox{
\subfigure[]{\includegraphics[width=0.95\columnwidth] {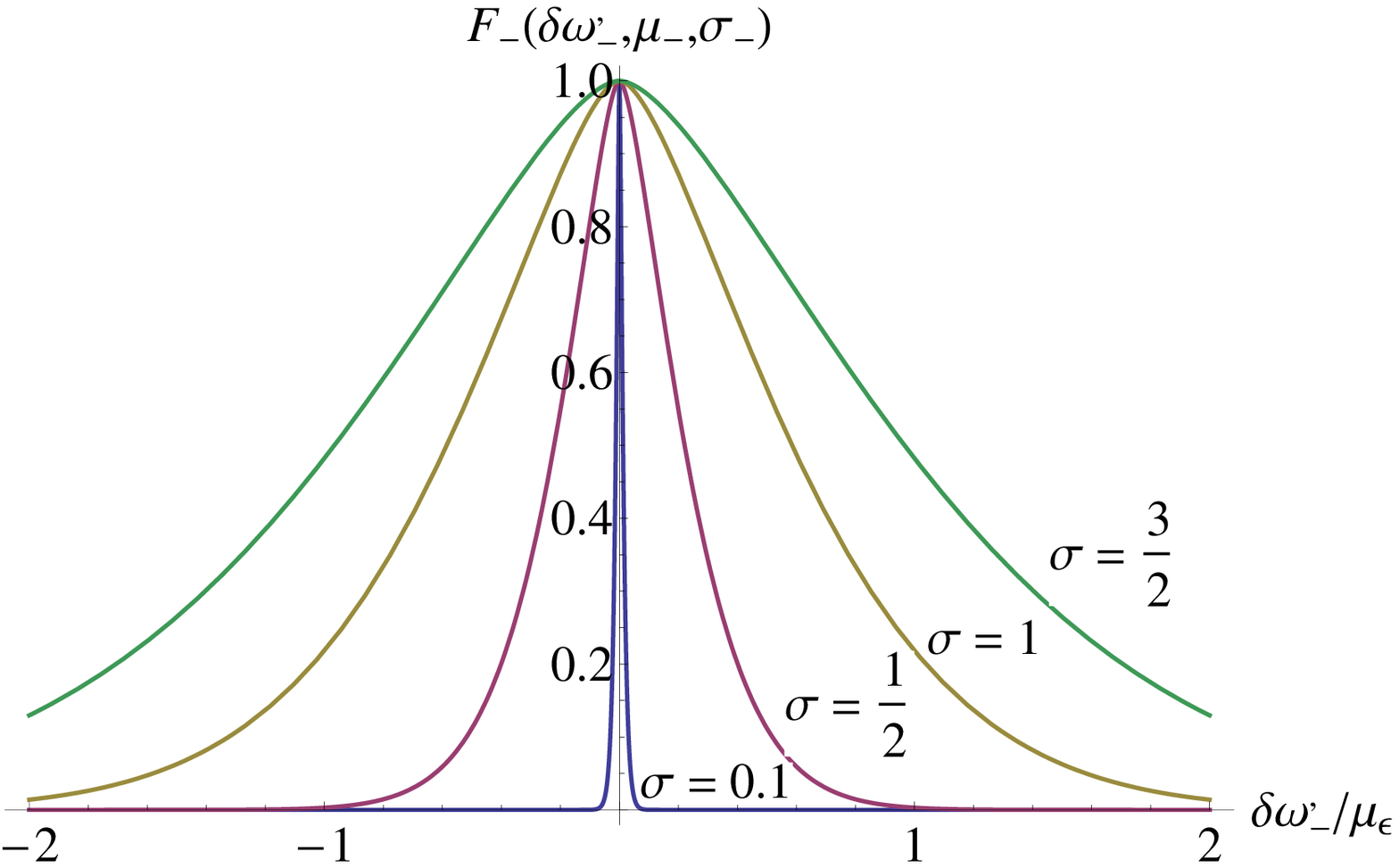}}
}
\caption{(color online) Plots of the distribution $F_-(\delta\omega_-^\prime,\mu_-,\sigma_-)$ that describes the contribution of static field inhomogeneities to the shape of the line that occurs in the FID experiment that is used to measure the $T_2^\ast$ time of the ground state spin. (a) the distribution for different values of $\mu = \mu_{{\cal B}}/\mu_{{\cal E}} = 0,1/2,1,5/4$  and $\sigma = \sigma_{{\cal B}}/\mu_{{\cal E}} = 0.1$. (b) the distribution for different values of $\sigma = 0.1,1/2,1,3/2$  and $\mu = 0$. Note that each distribution has been normalized such that its maximum is 1.}
\label{fig:Fdistribution}
\end{center}
\end{figure}

The distribution $F_-$ is plotted in Fig. \ref{fig:Fdistribution} for different values of the relative axial magnetic field mean $\mu$ and variance $\sigma$. The plots clearly demonstrate that the width of the $F_-$ distribution is highly dependent on $\mu$, such that it increases significantly for small increases in $\mu$ until $\mu \sim 1$. In the limit $\mu \gg 1$ it can be seen that the distribution has reached its maximum width and is very similar to a normal distribution. This is consistent with the distribution of ${\cal R}$ being dominated by the distribution of ${\cal B}_z$ when the mean axial magnetic field is much larger than the mean non-axial electric-strain field. Hence, it can be concluded that due to the dominance of magnetic inhomogeneities,  the contribution to the spin's $T_2^\ast$ time from statistical inhomogeneities in the static fields is dramatically reduced for field configurations where $\mu = \mu_{{\cal B}}/\mu_{{\cal E}} < 1$. This conclusion is consistent with the observations of the recent electric field sensing demonstration \cite{efield} and has significant implications for the spin's other sensing and QIP applications.

\section{Spin relaxation and dephasing}

As noted in section III, the ground state spin interacts with time-dependent incoherent electric and magnetic fields and crystal vibrations, which introduce the spin relaxation $\gamma_{ij}^r$ and dephasing $\gamma_{ij}^p$ rates into the evolution of the spin. The incoherent electric and magnetic fields arise from the thermal radiation field and  fluctuating magnetic and electric impurities and the crystal vibrations arise from the thermal motion of the crystal nuclei. Typical of spin systems, spontaneous radiative emission and contributions from the thermal field are negligible and can be safely ignored.\cite{EPR} As identified in Part I, the spin's interaction with electric fields is much smaller than its interactions with magnetic and strain fields. Consequently, the relaxation and dephasing rates are expected to be well described by just the spin's interactions with magnetic impurities and lattice vibrations. The contributions to relaxation and dephasing arising from interactions with magnetic impurities, including their dependence on the static magnetic field, has been studied in some detail for NV spins in both type Ib and type IIa diamond.\cite{highfield,dephasing1,dephasing2,dephasing3,dephasing4,dephasing5,dephasing6,dephasing7,dephasing8} The additional influence of electric-strain fields on these interactions has not yet been studied and this will be the subject of future work. In this section, the contributions to relaxation and dephasing arising from the spin's interaction with lattice vibrations will be theoretically developed for the first time.

The linear interaction of the spin with the vibrations of the crystal is described by the potential \cite{stoneham}
\begin{eqnarray}
\hat{V}_\mathrm{vib} = \sum_i\sum_{u,q}\left.\frac{\partial\hat{V}_{Ne}(\vec{r}_i,\vec{R})}{\partial Q_{u,E,q}}\right|_{\vec{R}_0}\sqrt{\frac{\hbar}{2\omega_{u,E}}}(\hat{b}_{u,E,q}+\hat{b}_{u,E,q}^\dagger) \nonumber \\
\end{eqnarray}
where as for the interaction of the spin with a static strain field, \cite{partI} $Q_{u,E,q}$ is defined as the $u^{th}$ mass-weighted normal displacement coordinate of symmetry $(E,q)$ that corresponds to an eigenmode of the ground triplet with frequency $\omega_{u,E}$ in the harmonic approximation, $\vec{R}_0$ are the ground state equilibrium coordinates, and $\hat{b}_{u,E,q}$ and $\hat{b}_{u,E,q}^\dagger$ are the vibration annihilation and creation operators respectively. Constructed in an analogous fashion to the matrix representation of the spin's interaction with the strain field $\hat{V}_\mathrm{str}$,\cite{partI} the matrix representation of $\hat{V}_\mathrm{vib}$ in the spin basis $\{\ket{0},\ket{-},\ket{+}\}$ is
\begin{widetext}
\begin{eqnarray}
&&\hat{V}_\mathrm{vib} = \sum_uq_u\sqrt{\frac{\hbar}{2\omega_{u,E}}}
\left[\left(\begin{array}{ccc}
0 & \frac{\chi}{\sqrt{2}}(e^{i\frac{\phi_{{\cal E}}}{2}}s_{\frac{\theta}{2}} - e^{-i\frac{\phi_{{\cal E}}}{2}}c_{\frac{\theta}{2}}) & \frac{\chi}{\sqrt{2}}(e^{i\frac{\phi_{{\cal E}}}{2}}c_{\frac{\theta}{2}} + e^{-i\frac{\phi_{{\cal E}}}{2}}s_{\frac{\theta}{2}}) \\
\frac{\chi}{\sqrt{2}}(e^{-i\frac{\phi_{{\cal E}}}{2}}s_{\frac{\theta}{2}} - e^{i\frac{\phi_{{\cal E}}}{2}}c_{\frac{\theta}{2}}) & -s_{\theta}c_{\phi_{{\cal E}}} & -c_{\theta}c_{\phi_{{\cal E}}} \\
\frac{\chi}{\sqrt{2}}(e^{-i\frac{\phi_{{\cal E}}}{2}}c_{\frac{\theta}{2}} + e^{i\frac{\phi_{{\cal E}}}{2}}s_{\frac{\theta}{2}}) & -c_{\theta}c_{\phi_{{\cal E}}} & s_{\theta}c_{\phi_{{\cal E}}}\\
\end{array}\right){\cal Q}_{u,E,x} \right.\nonumber \\
&&+\left.\left(\begin{array}{ccc}
0 & \frac{i\chi}{\sqrt{2}}(e^{i\frac{\phi_{{\cal E}}}{2}}s_{\frac{\theta}{2}} + e^{-i\frac{\phi_{{\cal E}}}{2}}c_{\frac{\theta}{2}}) & \frac{i\chi}{\sqrt{2}}(e^{i\frac{\phi_{{\cal E}}}{2}}c_{\frac{\theta}{2}} - e^{-i\frac{\phi_{{\cal E}}}{2}}s_{\frac{\theta}{2}}) \\
\frac{-i\chi}{\sqrt{2}}(e^{-i\frac{\phi_{{\cal E}}}{2}}s_{\frac{\theta}{2}} + e^{i\frac{\phi_{{\cal E}}}{2}}c_{\frac{\theta}{2}}) & -s_{\theta}s_{\phi_{{\cal E}}} & -c_{\theta}s_{\phi_{{\cal E}}} \\
\frac{-i\chi}{\sqrt{2}}(e^{-i\frac{\phi_{{\cal E}}}{2}}c_{\frac{\theta}{2}} - e^{i\frac{\phi_{{\cal E}}}{2}}s_{\frac{\theta}{2}}) & -c_{\theta}s_{\phi_{{\cal E}}} & s_{\theta}s_{\phi_{{\cal E}}}\\
\end{array}\right){\cal Q}_{u,E,y}\right] \nonumber \\
\end{eqnarray}
\end{widetext}
where $q_u = -\sqrt{2}s_{2,5}\rdm{a_1}{\partial\hat{V}_{Ne}/\partial Q_{u,E}|_{\vec{R}_0}}{e}$, $\chi = -\frac{\sqrt{2}s_{2,6}+s_{1,8}}{2\sqrt{2}s_{2,5}}$, $s_{n,m}$ are the spin coupling coefficients, $\rdm{a_1}{\ }{e}$ are the molecular orbital reduced matrix elements of the center (refer to Part I for further details), and ${\cal Q}_{u,E,q} = \hat{b}_{u,E,q}+\hat{b}_{u,E,q}^\dagger$.

Applying time-dependent perturbation theory, the above matrix representation can be used to derive the first order spin-lattice transition rates, which will contribute to both the relaxation and dephasing rates of the spin. The first order spin-lattice transition rates are
\begin{eqnarray}
&&W_{\pm\rightarrow0}^{\mathrm{vib}(1)} = \frac{\pi\chi^2}{\hbar}\frac{\overline{q^2}(\omega_{\pm})\rho_E(\omega_{\pm})}{\omega_{\pm}}
[n_T(\omega_{\pm})+1] \nonumber \\
&&W_{0\rightarrow\pm}^{\mathrm{vib}(1)} = \frac{\pi\chi^2}{\hbar}\frac{\overline{q^2}(\omega_{\pm})\rho_E(\omega_{\pm})}{\omega_{\pm}}
n_T(\omega_{\pm}) \nonumber
\end{eqnarray}
\begin{eqnarray}
&&W_{+\rightarrow-}^{\mathrm{vib}(1)} = \frac{\pi}{\hbar}\cos^2\theta\frac{\overline{q^2}(\omega_{+-})\rho_E(\omega_{+-})}{\omega_{+-}}
[n_T(\omega_{+-})+1] \nonumber \\
&&W_{-\rightarrow+}^{\mathrm{vib}(1)} = \frac{\pi}{\hbar}\cos^2\theta\frac{\overline{q^2}(\omega_{+-})\rho_E(\omega_{+-})}{\omega_{+-}}
n_T(\omega_{+-})
\end{eqnarray}
where $\overline{q^2}(\omega)$ is the average of $q_u^2$ over all $E$ symmetric vibrations of frequency $\omega$, $\rho_E(\omega)$ is the density of $E$ symmetric vibrations at frequency $\omega$, and $n_T(\omega) = 1/(e^{\hbar\omega/k_BT}-1)$ is the mean occupation number of thermal vibrations given by the Bose-Einstein factor.

The NV transition frequencies $\omega_\pm$ and $\omega_{+-}$ occupy the very low frequency end of the vibrational frequencies of diamond, which range from zero to the Debye frequency of diamond $\omega_D=38.76$ THz.\cite{debyefreq} In the low frequency limit, the long wavelength acoustic modes of the lattice have the well known Debye density $\rho_E(\omega)\approx\rho_a\omega^2$, where $\rho_a$ is a constant related to the acoustic velocity in diamond, and the electron-vibration interaction for the non-local acoustic modes has the approximate form $\overline{q^2}(\omega)\approx\omega \overline{q_a^2}$, where $\overline{q_a^2}$ is a constant.\cite{maradudin} Additionally, for temperatures $k_BT\gg\hbar\omega_{\pm}$, $\hbar\omega_{+-}$, the thermal occupations of the vibrational modes with frequencies corresponding to the NV transition frequencies can be approximated by $n_T(\omega)\approx k_BT/\hbar\omega$.\cite{stoneham} The first order transition rates for temperatures $k_BT\gg\hbar\omega_{\pm}$, $\hbar\omega_{+-}$ are thus approximately linear in $T$
\begin{eqnarray}
&&W_{\pm\rightarrow0}^{\mathrm{vib}(1)} \approx W_{0\rightarrow\pm}^{\mathrm{vib}(1)} \approx {\cal A}_a\chi^2\omega_{\pm}^2T \nonumber \\
&&W_{+\rightarrow-}^{\mathrm{vib}(1)} \approx W_{-\rightarrow+}^{\mathrm{vib}(1)} \approx {\cal A}_a\cos^2\theta\omega_{+-}^2
T
\end{eqnarray}
where ${\cal A}_a=\pi k_B\rho_a\overline{q_a^2}/\hbar^2$. Noting that for an axially aligned magnetic field, $\omega_{+-}^2\cos^2\theta=4{\cal B}_z^2$ and $\omega_\pm=({\cal D}\pm{\cal R})^2$, it is clear that the first order transitions between different spin states depend differently on the field parameters.

Since the electron-vibration interaction $\overline{q^2}(\omega)$ and the density of vibrational modes $\rho_E(\omega)$ increase at higher vibrational frequencies, the first order transitions will only be the dominant spin-lattice mechanisms at low temperatures, where there is only appreciable occupations of the low frequency vibrational modes. At higher temperatures, the occupation of the more numerous and strongly interacting higher frequency modes ensures that elastic and inelastic Raman scattering of vibrations will become the dominant spin-lattice mechanisms.\cite{stoneham} The inelastic scatterings will contribute to both relaxation and dephasing, whereas the elastic scatterings will contribute to just dephasing. Note that as the two-vibration absorption/ emission transitions involve vibrations of even lower frequencies than the first order transitions, they will be negligible at all temperatures. Expecting the most significant contributions to be from vibrational modes with frequencies $\omega\gg\omega_\pm$, $\omega_{+-}$, the elastic and inelastic Raman scattering rates are approximately
\begin{eqnarray}
&&W_{\pm\rightarrow0}^{\mathrm{vib}(2)} \approx W_{0\rightarrow\pm}^{\mathrm{vib}(2)} \approx W_{\pm\rightarrow\mp}^{\mathrm{vib}(2)} \approx \frac{1}{2}W_{0\rightarrow0}^{\mathrm{vib}(2)} \nonumber \\
&&\approx \frac{\pi\chi^2}{\hbar^2}\int_0^{\omega_D}\frac{\overline{q^2}^2(\omega)\rho_E^2(\omega)}{\omega^4}
n_T(\omega)[n_T(\omega)+1]d\omega \nonumber \\
&&W_{\pm\rightarrow\pm}^{\mathrm{vib}(2)}  \nonumber \\
&&\approx \frac{\pi}{\hbar^2}[1+\chi^2(1\pm\sin\theta\cos3\phi_{{\cal E}})+\chi^4] \nonumber \\
&&\int_0^{\omega_D}\frac{\overline{q^2}^2(\omega)\rho_E^2(\omega)}{\omega^4}
n_T(\omega)[n_T(\omega)+1]d\omega
\end{eqnarray}

\begin{figure}[hbtp]
\begin{center}
\mbox{
\subfigure[]{\includegraphics[width=0.85\columnwidth] {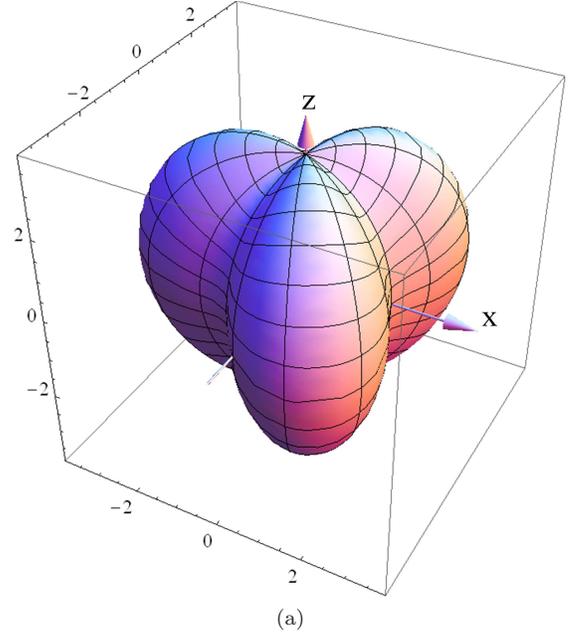}}}
\mbox{
\subfigure[]{\includegraphics[width=0.8\columnwidth] {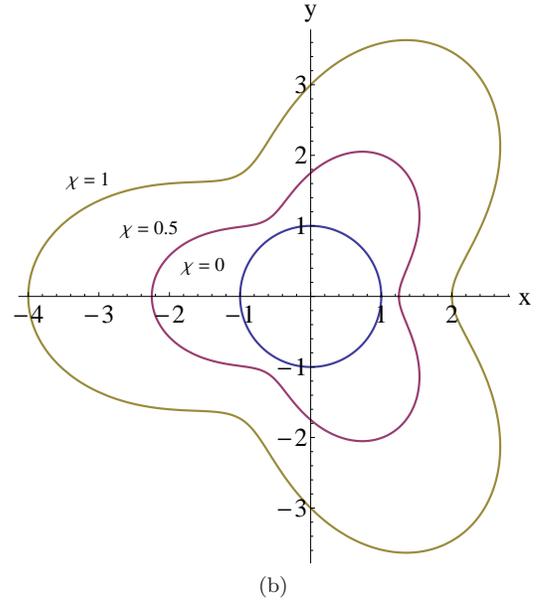}}
}
\caption{(color online) Plots of the dimensionless elastic vibration scattering parameter $[1+\chi^2(1-\sin\theta\cos3\phi_{{\cal E}})+\chi^4]$ as: (a) a function of the azimuthal $0\leq\phi_{{\cal E}}\leq2\pi$ and polar $0\leq\theta\leq\pi$ angles for $\chi = 1$; and (b) as a function of $\phi_{{\cal E}}$ for $\theta = \frac{\pi}{2}$ and different values of $\chi$ as indicated. Coordinate axes are provided for reference to figure \ref{fig:centre} where the field angles $\phi_{{\cal E}}$ and $\theta$ are defined.}
\label{fig:scatteringplots}
\end{center}
\end{figure}

The dependence of the elastic scattering rates of the $\ket{\pm}$ spin states on the static field angles is particularly interesting. Figure \ref{fig:scatteringplots} contains polar plots of the dimensionless scattering parameter $[1+\chi^2(1\pm\sin\theta\cos3\phi_{{\cal E}})+\chi^4]$ as a function of the $\theta$ and $\phi_{{\cal E}}$ field angles for different values of $\chi$. As can be seen, for $|\chi|  >0$ the elastic scattering rates are minimum at $\phi_{{\cal E}} = 0, \frac{2\pi}{3}, \frac{4\pi}{3}$, mimicking the structural symmetry of the defect center. Hence, it appears possible for the orientation of the non-axial electric-strain field to be tuned so that the elastic scattering rates are minimized or maximized with the difference in the minimum and maximum rates $2\chi^2$ determined by the ratio of spin coupling coefficients $\chi$. Note that $\chi$ is currently unknown.

Considering temperatures not so high that optical modes are appreciably occupied in thermal equilibrium, there will be two distinct contributions to the integrals in the Raman scattering rates. The first will be from the acoustic modes which have electron-vibration interaction $\overline{q^2}(\omega)\approx\omega\overline{q_a^2}$ and mode density $\rho_E(\omega)\approx\rho_a\omega^2$. The second will be from the strongly interacting local modes of the NV center which have frequencies $\omega_l\sim65$ meV. \cite{hamer} The contribution from the local modes can be represented by the electron-vibration interaction $\overline{q^2}(\omega_l)\approx\overline{q_l^2}$ and a sharp peak in the density of modes centered at $\rho_E(\omega_l)=\rho_l$ with width $\sigma_l$. Given the separable contributions, the integrals can be evaluated and the Raman scattering rates reduce to
\begin{eqnarray}
&&W_{\pm\rightarrow0}^{\mathrm{vib}(2)} \approx W_{0\rightarrow\pm}^{\mathrm{vib}(2)} \approx W_{\pm\rightarrow\mp}^{\mathrm{vib}(2)} \approx \frac{1}{2}W_{0\rightarrow0}^{\mathrm{vib}(2)} \nonumber \\
&&\approx \chi^2\left[{\cal A}_l^2
n_T(\omega_l)[n_T(\omega_l)+1]+{\cal A}_a^2\frac{4\pi^3k_B^3}{15\hbar^3}T^5\right] \nonumber \\
&&W_{\pm\rightarrow\pm}^{\mathrm{vib}(2)}  \nonumber \\
&&\approx [1+\chi^2(1\pm\sin\theta\cos3\phi_{{\cal E}})+\chi^4] \nonumber \\
&&\left[{\cal A}_l^2
n_T(\omega_l)[n_T(\omega_l)+1]+{\cal A}_a^2\frac{4\pi^3k_B^3}{15\hbar^3}T^5\right]
\end{eqnarray}
where ${\cal A}_l^2=\pi\overline{q_l^2}^2\rho_l^2\sigma_l^2/\hbar^2\omega_l^4$.
Note that the integral over the acoustic modes was evaluated in the limit $\omega_D\rightarrow\infty$ in order to obtain the simple $T^5$ factor. Given the temperatures being considered, for which high frequency optical modes are not appreciably occupied, this extension of the integral is expected to be inconsequential. Hence, the Raman scattering rates depend on temperature in two distinct ways due to the distinct contributions of a few strongly interacting local modes and many weakly interacting non-local acoustic modes.

Combining the magnetic ($W^B$) and spin-lattice ($W^{\mathrm{vib}(1)}$ and $W^{\mathrm{vib}(2)}$) contributions, the relaxation and dephasing rates of the ground state spin are
\begin{eqnarray}
\gamma_{\pm0}^r &=& W_{\pm\rightarrow0}^{B}+W_{\pm\rightarrow0}^{\mathrm{vib}(1)}+W_{\pm\rightarrow0}^{\mathrm{vib}(2)} \notag \\
\gamma_{0\pm}^r &=&
W_{0\rightarrow\pm}^{B}+W_{0\rightarrow\pm}^{\mathrm{vib}(1)}+W_{0\rightarrow\pm}^{\mathrm{vib}(2)} \notag \\
\gamma_{\pm\mp}^r &=&
W_{\pm\rightarrow\mp}^{B}+W_{\pm\rightarrow\mp}^{\mathrm{vib}(1)}+W_{\pm\rightarrow\mp}^{\mathrm{vib}(2)}\notag \\
\gamma_{\pm0}^p &=& \frac{1}{2}(\gamma_{\pm0}^r+\gamma_{0\pm}^r)+W_{0\rightarrow0}^{B}+W_{\pm\rightarrow\pm}^{B}+W_{0\rightarrow0}^{\mathrm{vib}(2)}\notag\\
&&+W_{\pm\rightarrow\pm}^{\mathrm{vib}(2)} \notag \\
\gamma_{+-}^p &=& \frac{1}{2}(\gamma_{+-}^r+\gamma_{-+}^r)+W_{-\rightarrow-}^{B}+W_{+\rightarrow+}^{B}+W_{-\rightarrow-}^{\mathrm{vib}(2)} \notag \\
&&+W_{+\rightarrow+}^{\mathrm{vib}(2)}
\label{eq:dephasingrates}
\end{eqnarray}
As described elsewhere, the magnetic contributions are highly dependent on the static magnetic field,\cite{dephasing1,dephasing2,dephasing5,dephasing6,dephasing8} but are essentially temperature independent for temperatures $>20$ K, due to the impurity spins easily reaching equal Boltzmann populations of their spin sub-levels at low temperatures.\cite{highfield} The spin-lattice contributions are instead weakly dependent on the static fields, but have distinct functions of temperature that arise from different interactions with lattice vibrations.

For a simple ODMR experiment using the $\omega_-$ transition, the spin relaxation $T_1$ and dephasing $T_2$ times are defined by $1/T_1 = \gamma_{-0}^r+\gamma_{0-}^r$ and $1/T_2=\gamma_{-0}^p$, which using the above are explicitly
\begin{eqnarray}
\frac{1}{T_1}
&\approx& 2\Gamma_{B1}+2\Gamma_{\mathrm{vib}1}\omega_-^2T+2\Gamma_{\mathrm{vib}2}n_T(\omega_l)[n_T(\omega_l)+1]\notag\\
&&+2\Gamma_{\mathrm{vib}3}T^5\notag\\
\frac{1}{T_2}
&\approx&\frac{1}{2T_1}+\Gamma_{B2}+\left[\frac{1}{\chi^2}+(3-\sin\theta\cos3\phi_{{\cal E}})+\chi^2\right]\notag\\
&&\left[\Gamma_{\mathrm{vib}2}n_T(\omega_l)[n_T(\omega_l)+1]+\Gamma_{\mathrm{vib}3}T^5\right]\notag
\end{eqnarray}
where $\Gamma_{\mathrm{vib}1}=\chi^2{\cal A}_a$, $\Gamma_{\mathrm{vib}2}=\chi^2{\cal A}_l^2$, and $\Gamma_{\mathrm{vib}3}=4\pi\chi^2k_B^3{\cal A}_a^2/15\hbar^3$ are constants that are independent of the static fields and temperature, and $\Gamma_{B1}$ and $\Gamma_{B2}$ are the magnetic contributions that are dependent on the static fields, but effectively temperature independent. An analogous expression can be simply derived for the $\omega_+$ transition.

Noting that $n_T(\omega_l)[n_T(\omega_l)+1]\approx n_T(\omega_l)$ for $k_BT<\hbar\omega_l$, the contribution to $1/T_1$ from inelastic Raman scatterings of strongly interacting local modes has been experimentally observed.\cite{relaxtemp} Likewise, the $T$ and $T^5$ contributions from the weakly interacting acoustic modes have also been observed.\cite{highfield} The spin-lattice contribution to $1/T_1$ only depends on the static fields through the presence of $\omega_-^2$ in the linear temperature term. This dependence on the static fields has not yet been observed, which is most likely due to the insignificance of the linear term at ambient temperatures and the fact that most previous measurements have been performed using NV ensembles, where resonant interactions between NV sub-ensembles and between NV centers and P1 centers at magnetic fields around ${\cal B}\sim$ 0, 0.96, 1.44 1.68 GHz,\cite{dephasing1,dephasing2} will most likely have masked the relatively weak dependence of the linear term. The dephasing rate $1/T_2$ is dominated by the contributions from magnetic interactions with little observed temperature dependence in small static fields.\cite{dephasing7}  Since the magnetic contribution is governed by the impurity concentration,\cite{dephasing3} it may be possible to observe the spin-lattice contribution in highly pure samples. It would indeed be interesting to observe the tuning of $1/T_2$ using an electric-strain field via the spin-lattice elastic scattering parameter $[1/\chi^2+(3-\sin\theta\cos3\phi_{{\cal E}})+\chi^2]$.

\section{Conclusion}

In this article, a general solution was obtained for the NV spin in any given electric-magnetic-strain field configuration for the first time, and the influence of the fields on the evolution of the spin was examined. In particular, the control of the spin's susceptibility to inhomogeneities in the static fields and crystal vibrations was examined in detail. The analysis of the effects of inhomogeneous fields revealed the field configurations required to switch between different noise dominated regimes. The analysis of the spin's interactions with crystal vibrations yielded observable effects that are consistent with previous observations and also the basis for future investigations into the potential tuning of the spin's dephasing rate. Hence, this work has provided essential theoretical tools for the precise control and modeling of this remarkable spin in its current and future quantum metrology and QIP applications.

\begin{acknowledgments}
This work was supported by the Australian Research Council under the
Discovery Project scheme (DP0986635 and DP0772931), the EU commission (ERC grant SQUTEC), Specific Targeted Research Project  DIAMANT and the integrated project SOLID. F.D. wishes to acknowledge the Baden-Wuerttemberg Stiftung Internat. Spitzenforschung II MRI.
\end{acknowledgments}

\end{document}